\documentclass[10pt,aps,prb,twocolumn,superscriptaddress,showpacs]{revtex4-2}
\usepackage{amsmath}
\usepackage{amsfonts}
\usepackage{graphicx} % Include figure files
\usepackage{float}
\usepackage{dcolumn}
\usepackage{bm}
\usepackage{xcolor}
\usepackage{subfigure}
\usepackage{cancel}

\newcommand{\avgw}{{\avleft w\avright}}
\newcommand{\argu}[1]{\!\left({#1}\right)}

\newcommand{\varphivec}{\mbox{\boldmath$\varphi$}}
\newcommand{\etavec}{\mbox{\boldmath$\eta$}}

\newcommand{\be}{\begin{equation}}
\newcommand{\ee}{\end{equation}}
\newcommand{\avleft}{{\langle\langle \,}}
\newcommand{\avright}{\,\rangle \rangle}
\begin{document}

\newcommand{\mk}[1]{ {\color{blue} {#1}} }

\frenchspacing

\title{Superdiffusive quantum work and adiabatic quantum evolution in finite temperature chaotic Fermi systems}
%\title{Quantum work statistics in finite temperature disordered Fermi liquids}
%Statistical behavior of chaotic fermionic nanosystems at finite temperatures}}

\author{Andr\'as Grabarits}
\affiliation{Department of Theoretical Physics, Institute of Physics, 
Budapest University of Technology and Economics, M\H uegyetem rkp. 3., H-1111 Budapest, Hungary}
\affiliation{MTA-BME Quantum Dynamics and Correlations Research Group, 
Budapest University of Technology and Economics,  M\H uegyetem rkp. 3., H-1111 Budapest, Hungary}
%Budapest University of Technology and Economics, Institute of Physics, 
%Budafoki \'ut 8., H-1111 Budapest, Hungary}
\author{M\' arton Kormos}
\affiliation{Department of Theoretical Physics, Institute of Physics, 
Budapest University of Technology and Economics, M\H uegyetem rkp. 3., H-1111 Budapest, Hungary}
\affiliation{MTA-BME Quantum Dynamics and Correlations Research Group, 
Budapest University of Technology and Economics,  M\H uegyetem rkp. 3., H-1111 Budapest, Hungary}
\author{Izabella Lovas}
\affiliation{Kavli Institute for Theoretical Physics, University of California, Santa Barbara, CA 93106, USA}
\author{Gergely Zar\'and}
\affiliation{Department of Theoretical Physics, Institute of Physics, 
Budapest University of Technology and Economics, M\H uegyetem rkp. 3., H-1111 Budapest, Hungary}
\affiliation{MTA-BME Quantum Dynamics and Correlations Research Group, 
Budapest University of Technology and Economics,  M\H uegyetem rkp. 3., H-1111 Budapest, Hungary}
%\affiliation{BME-MTA Exotic Quantum Phases `Lend\"ulet' Research Group, Institute of Physics, 
%Budapest University of Technology and Economics, Budafoki \'ut 8., H-1111 Budapest, Hungary}

\begin{abstract}
We  study the full distribution of quantum work in generic, noninteracting, disordered fermionic nanosystems at finite temperature. 
We derive an analytical  determinant formula for the characteristic function of work statistics for quantum quenches starting from a thermal initial state.
For work small compared to the thermal energy of the Fermi gas,  work distribution is Gaussian, and the variance of work is proportional 
to the average work, while in the low temperature or large work limit, a non-Gaussian distribution with 
superdiffusive work fluctuations is observed. 
Similarly, the time dependence of the probability of adiabaticity crosses over from an exponential to a stretched exponential 
behavior. For large enough average work, the work distribution becomes universal, and depends only on the temperature 
and the mean work.  Apart from initial low temperature transients, work statistics are well-captured by a Markovian energy-space diffusion process of hardcore particles, starting from a thermal  initial state. 
Our findings can be verified by  measurements on nanoscale circuits or via single qubit interferometry.
\end{abstract}

\maketitle

\section{Introduction}

Even though the change of energy and the related concepts of heat and work are fundamental quantities in thermodynamics, their understanding in the context of quantum systems is still incomplete, and only well-established in some limiting cases, such as within the quasi-stationary approximation or near equilibrium situations. Generalizations to arbitrary non-equilibrium processes in quantum systems raise highly nontrivial unsettled questions. 

In the quantum mechanical context, `work' $W$ is commonly defined as energy transfer during some deformation of the system. As a consequence, it requires a two-time measurement scheme~\cite{Hanggi,workreview}, in which first the energy $E_\text{i}(0)$ of the initial state is measured at time $t = 0$, and then, in a second measurement at time $t$,
the energy $E_\text{f}(t)$ of the time evolved system is determined, where $E_\text{f}(t)$ is an energy eigenvalue of the final Hamiltonian $\hat H(t)= \hat H_\text{f}$. The total absorbed energy,
$E_\text{f}(t) - E_\text{i}(0)$, can be split into two parts: an adiabatic contribution corresponding to the essentially trivial shifts of the energy levels, and a nontrivial part, which we define as \emph{work}, that captures the contribution of non-adiabatic particle-hole excitations. 

Work thus naturally  becomes a \emph{statistical}  quantity due to thermal and quantum fluctuations, and its 
 proper description is provided by its full distribution function,
giving the probabilities of energy transferred to ($W>0$)
or extracted from ($W<0$) the system at a given time $t$ and at temperature $T$.  Recent experimental developments made it possible to 
measure work distribution and  to  verify 
 non-equilibrium fluctuation relations~\cite{jarz,crooks}  in quantum 
 systems ranging from individual molecules~\cite{bioreview,molecule1,molecule2} through nuclear spins 
 subject to magnetic fields~\cite{batalhao}, to cold atoms~\cite{cerisola} and to mesoscopic 
 grains~\cite{pekola,pekola2}.

Here we focus on small disordered Fermi liquid systems such as small metallic grains, 
which  can be described in terms of non-interacting fermions, 
and we study their work statistics at finite temperatures.  Our work is motivated by recent experimental developments in the field 
of  driven nanoscale circuits~\cite{pekola,pekola2}, which allow one to investigate thermodynamics and non-equilibrium processes
 in the quantum realm. Intriguing theoretical proposals also exist to extract quantum work distributions 
and Loschmidt echoes in such systems, by performing measurements on ancilla qubits, coupled 
to the physical system studied~\cite{Mazzola,Dorner,SquidLuttinger}. 

Work statistics have been studied 
theoretically in clean systems such as spin chains~\cite{Dorosz,Silva,Smacchia}, Luttinger liquids~\cite{Dora}, near critical systems and field theories~\cite{Gambassi,Smacchia,Spyros,Palmai}, noninteracting~\cite{Yi1,Yi2} and interacting bosons~\cite{Perfetto}. Quantum corrections and small 
work expansions have also been derived~\cite{Quan1,Quan2}.
Less attention has been devoted to 
  \emph{randomness}, however, although it  plays a key role  in chaotic mesoscopic and nanoscale systems.
While the average energy absorption has been investigated in chaotic and disordered systems quite some time ago 
in the pioneering works of Wilkinson~\cite{wilkinsondiff1,wilkinson3,wilkinson4,wilkinson7} and 
Kravtsov~\cite{Kravtsov1,Kravtsov2,Kravtsov3},
the full distribution, $P(W,\,t)$, has started to attract attention much more recently.  Most of the works on disordered systems have focused on  sudden quenches~\cite{Garcia-Mata,Lobejko,Arrais1,Arrais2,delCampo1,delCampo2},  while the first results on generic quench protocols in driven fermionic random systems considered only the zero temperature limit~\cite{Grabarits1,Grabarits2}.

Building on the results presented in Refs.~\onlinecite{Grabarits1,Grabarits2}, here we explore the work statistics of chaotic Fermi systems at finite temperatures. Following Refs.~\onlinecite{Grabarits1,Grabarits2},  we define work in the following way. We separate the total energy absorption, $E_\text{f}(t)-E_\text{i}(0)$, into two parts. The first contribution, $E_ i(t)-E_\text{i}(0)$, accounts for the absorbed or emitted energy in
adiabatic processes, where the energy levels of $\hat H(t)$ are shifted but their occupations do not change during the quantum quench. The second contribution to the energy absorption, $E_\text{f}(t) - E_\text{i}(t)$, accounts for non-adiabatic particle-hole excitations created during the quench protocol. We thus define work
as the energy of the final state with respect to the \emph{adiabatically evolved} 
initial state, $E_\text{i}\to E_\text{i}(t)$,
\be
W \equiv E_\text{f}(t) - E_\text{i}(t)\;.
\ee
This definition allows us to separate  clearly  adiabatic ($W=0$) and non-adiabatic ($W\ne0$) processes, as
\be 
P(W,\,t) = P^\text{ad}(t) \,\delta(W) + P^{\text{reg}}(W,\,t)\;,
\label{P_t(w)_structure}
\ee
where $P^\text{ad}(t)$ is the probability of adiabatic evolution characterized by the final energy level populations being equal to the initial ones, and $P^{\text{reg}}(W,\,t)$ is a regular part.
In Refs.~\onlinecite{Grabarits1,Grabarits2}, we have shown at $T=0$ temperature  that,  in chaotic 
Fermi systems, described by random matrix theory,
$P^\text{ad}_{T=0}(t)$ decays as a stretched exponential, and we have derived approximate 
expressions for  the non-Gaussian regular part, $P^{\text{reg}}_{T=0}(W,\,t)$, using bosonization 
and mean field methods. 
Work, in particular,  has been shown to  display a superdiffusive behavior~\cite{Grabarits1,Grabarits2}, 
and the work statistics has been found to exhibit  interesting universal features at $T=0$ temperature. 

Real systems are, however, always subject to finite temperature. 
Here we therefore extend our previous, $T=0$ temperature fermionic random matrix approach~\cite{Grabarits1,Grabarits2} 
 to \emph{finite} temperatures. The random matrix models studied here capture the
 properties of  driven chaotic and disordered Fermi-systems  such as metallic nano-grains~\cite{matrixreview}.
The latter represent prototypical examples  of chaotic Fermi systems, and
 form  an ideal experimental  testbed to study the interplay  
of  many-body time evolution  and disorder~\cite{Kastnerreview,Coulomb_book,pekola,pekola2,calorimetry1}.

We generalize the  determinant formula 
 of Ref.~\onlinecite{Grabarits1} to $ T\ne 0 $ temperatures, where the quantum mechanical average is taken with respect to a thermal density matrix as opposed to the ground state (i.e., the initial state is drawn from a thermal distribution)\cite{footnoteA}.
We compute the full work distribution as well as its variance and expectation value.  
We find that two main regimes must be 
distinguished at finite temperature: 
for average works $\avleft W\avright$ small compared to the total thermal energy of the Fermi liquid, $E_T\sim T^2$, 
work distribution is close to Gaussian, and its variance  is proportional to $\delta W^2\sim \avleft W\avright$, 
as naively and classically expected, based upon the central limit theorem. 
For injected work larger than $E_T$, however,  
work distribution {\em deviates} from Gaussian, and work is superdiffusive, 
$\delta W^2\sim \avleft W\avright^{3/2}$, as found at $T=0$ temperature.

As mentioned before, our definition of work allows us to study the probability of adiabatic 
processes even at finite temperatures. We find that $P^\text{ad}(t)  $ also displays markedly 
different behavior in the small- and large work regimes.  

We also investigate finite temperature quantum work distribution from the perspective of \emph{universality}. 
Universality is a central concept in many areas of physics: it  refers to the independence
of certain properties (exponents, distributions etc.) from microscopic details, and the crucial role of 
symmetries and dimensionality~\cite{Cardy,Sachdev}.  Universal properties during non-equilibrium processes 
have been investigated in 
the context of non-thermal fixed points~\cite{Gasenzer,boseuniversality,boseuniversality2}, or
spin transport processes in strongly interacting Fermi gases~\cite{hydro1,hydro2}, 
and in relation to work statistics in quenched clean systems~\cite{Gambassi,Smacchia,Spyros}.

In the context of work statistics in disordered nano-grains, an even stronger sense of universality 
appears to emerge~\cite{Grabarits2}. For sufficiently large injected work, the work statistics is independent not only of
 such microscopic details as the  metallic grains' level spacing, $\delta\epsilon$, 
but even of the symmetry 
of the underlying Hamiltonian (unitary, orthogonal, or symplectic). As we demonstrate, 
the  universal behavior found at $T=0$ temperature carries over to finite temperatures too, and is well captured by the simple classical  `ladder' model constructed in Ref.~\onlinecite{Grabarits2}.

The paper is organized as follows. After setting up the model and the formalism in Sec.~\ref{sec:rev},  we review our earlier results on the zero temperature case in Sec.~\ref{sec:T0}. Sec.~\ref{sec:T>0} contains our quantum mechanical results on work statistics 
and the generalized determinant formula.
Section \ref{sec:SEP} is dedicated to the analytical results, including large and small work expansions, obtained from a classical Markovian diffusion process in energy space. We give our conclusions in Sec.~\ref{sec:concl}. 
 Details of the calculations can be found in the Appendices.

\section{Theoretical framework}
\label{sec:rev}

\begin{figure}[t!]
\includegraphics[width=0.8\columnwidth, trim={0 5cm 0 5cm} ]{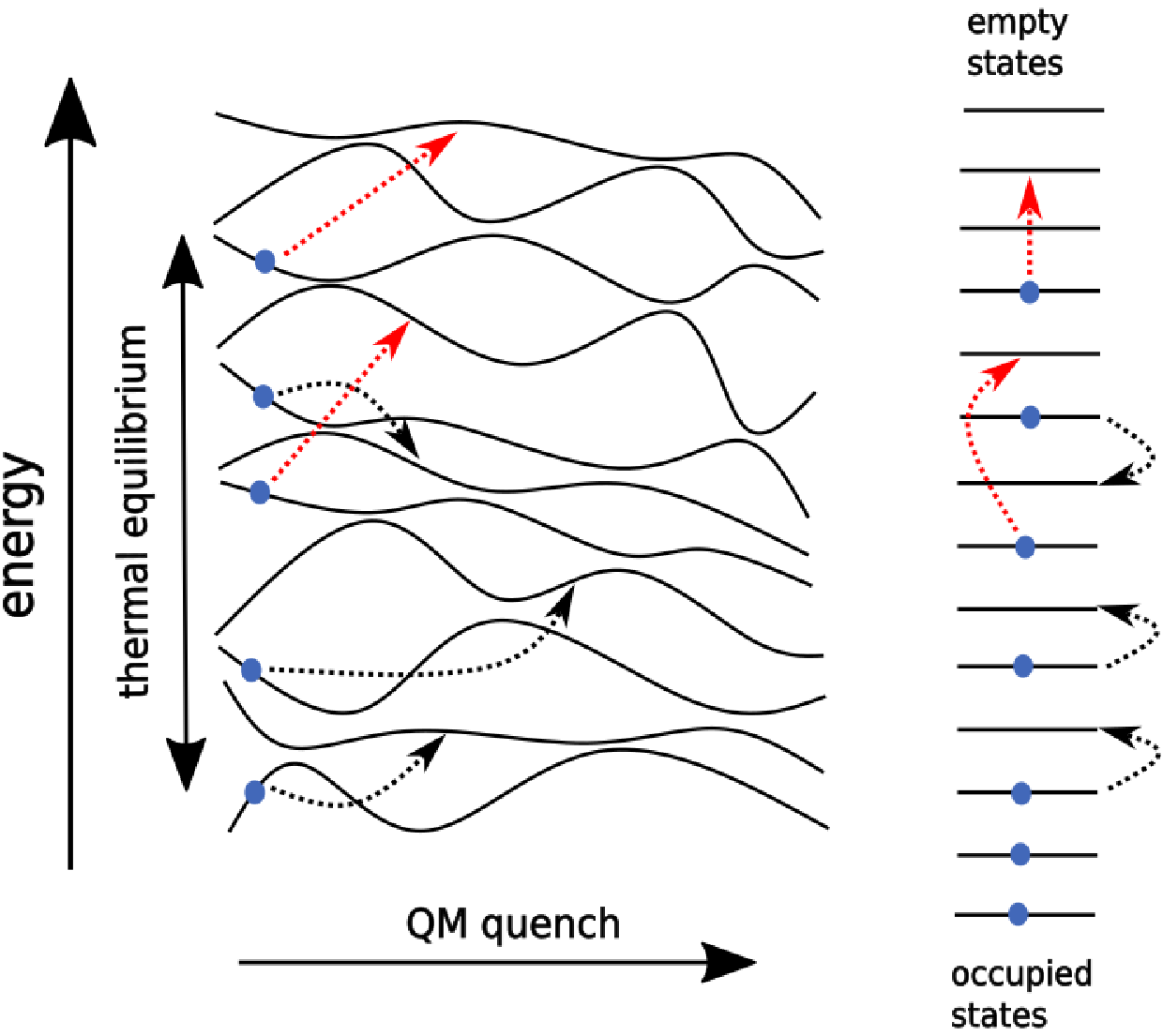}
\caption{\label{fig:sketch}
Left: Energy levels of a disordered nanograin during a quantum quench.  
 External time dependent gate voltages and magnetic fields induce quantum mechanical 
 level-to-level transitions and generate work. A thermal average is performed over initial states.
Right: Ladder model.  Particles move diffusively between uniformly spaced 
energy levels.  Initial, finite $T$ configurations are generated with appropriate 
   Boltzmann weights.}
\end{figure}

We model the dynamics of $M$ non-interacting fermions in generic disordered, metallic grains by the Hamiltonian
\begin{equation}\label{eq:H}
\hat{H}(t)=\sum_{i,j=1}^N \mathcal{H}_{ij}(t)\,\hat{a}_i^\dagger\,\hat{a}_j\,,
\end{equation}
where $\hat{a}_i^\dagger$ and $\hat{a}_i$ denote fermionic creation and annihilation operators satisfying $M=\sum_i \hat{a}_i^\dagger\,\hat{a}_i$, and $\mathcal{H}(t)$ is an $N\times N$ random matrix. In experiments, the grain is subject to time-dependent external gate voltages and magnetic fields~\cite{mesobook},
 leading to the movement and collisions of energy levels (avoided level crossings). These avoided level crossings  
 change the levels' occupations and thus the total energy of the system (see Fig.~\ref{fig:sketch}). 
We model time-dependent driving by the Hamiltonian~\cite{wilkinson2,wilkinson7,Grabarits1},
\begin{equation}\label{eq:protocol}
\mathcal{H}(t)=\mathcal{H}_1\cos\lambda(t)+\mathcal{H}_2\sin\lambda(t)
\end{equation}
with $\mathcal{H}_1$ and $\mathcal{H}_2$  independent $N\times N$ random matrices, generated 
with probabilities
\begin{equation}\label{eq:ph}
\mathcal{P}_\beta\left(\mathcal{H}\right)\sim e^{-\frac{\beta N}{4J^2}{\rm Tr}\mathcal{H}^2}
%/\mathcal{N_\beta}
\end{equation}
from the Gaussian orthogonal (GOE, $\beta=1$), unitary (GUE, $\beta=2$), or symplectic (GSE, $\beta=4$) ensembles.
These matrix ensembles describe disordered systems without and with spin in the presence of time-reversal symmetry (GOE and GSE, respectively) and systems without time-reversal symmetry (GUE)~\cite{matrixreview,RMreview,wilkinson6}.
 %$\mathcal{N}_\beta$ is a normalization constant and 
The energy scale $J$, set to  $J=1$ hereafter, fixes the bandwidth of the spectrum of $\mathcal{H}$. 
The form of Eq. \eqref{eq:protocol} ensures that we remain in the respective matrix 
 ensemble  throughout the whole quench process. Accordingly, at any instance, 
 energy levels display level repulsion, and the distribution of the spacing  $s$ 
between neighboring levels, $P_\mathrm{level,\beta}$, obeys Wigner-Dyson statistics~\cite{matrixreview,RMreview}, 
implying
$P_\mathrm{level,\beta}(s)\sim s^{\,\beta}$ for $s\approx 0$.

In this work,  we focus on  a uniform time 
evolution between $\mathcal H_1$ and $\mathcal H_2$ in the space of random matrices,
\begin{equation}
	\lambda(t)=v\,t,\,\,\,\,\,\,0\leq t\leq\tau\,.
\end{equation}
The velocity   $v$ here is related  to the average  velocity  in  the
space of random matrices as
\begin{equation}
	 v =\frac{1}{\sqrt NJ}\Big\langle \frac{\mathrm ds}{\mathrm dt}\Big\rangle_{\mathrm{RM}} \,,
\end{equation}
where $\mathrm ds$  denotes the arc length, 
$\mathrm ds^2=\mathrm{Tr}(\mathrm d\mathcal H\,\mathrm d\mathcal H)$, and 
$\langle\dots\rangle_\mathrm{RM}$  the averaging over the random matrix ensemble.

To characterize work, it is useful to define the instantaneous single particle eigenvectors $\etavec_{k,t}$ of 
${\cal H}(t)$,  and the corresponding creation operators,  $\hat b^\dagger_{k,t}$.
The many-body Hamiltonian can then be written in terms of these as
\be
\hat H(t)=\sum_{k=1}^N\varepsilon_k(t)\,\hat b^\dagger_{k,t}\hat b_{k,t}
\ee 
with the $\varepsilon_k(t)$'s referring to 
the instantaneous single particle energies. 

The average work can then be expressed in terms of the instantaneous occupations, $ f_k(t)  
\equiv \langle  \hat b^\dagger_{k,t}\hat b_{k,t} \rangle=\mathrm{Tr}\left[\hat b^\dagger_{k,t}\hat b_{k,t}\rho_0\right]$, as 
\be \avleft  W\avright 
=\sum_{k=1}^N\langle\varepsilon_k(t)\, (  	f_k(t)-f_k^{\,0} )\rangle^{\phantom{N}}_{\mathrm{RM}}\;,
\label{eq:Waver}	
\ee
where the double average $\avleft \dots\avright$ refers to a thermal average over the initial states containing $M$-particles, as well
as over the random matrix ensemble, and $f_k^{\,0} = f_k(0) = 
\langle\hat b^\dagger_{k,0}\hat b_{k,0}\rangle$. 
At $T=0$ temperature, the thermal average is replaced by a quantum average 
over the initial (ground) state of the fermions, while $f_k^{\,0,T=0}=\Theta\left(\varepsilon_\text{F}-\varepsilon_k^0\right)$, with $\Theta$ denoting the Heaviside function, $\varepsilon_k^0\equiv\varepsilon_k(0)$, and $\varepsilon_\text{F}$ the Fermi energy. 

Notice that in Eq.~\eqref{eq:Waver}, we define work somewhat differently than usual:
we subtract from the instantaneous energy of a given state the energy of 
the \emph{adiabatically evolved} state, $\sum_k \varepsilon_k(t)\,f_k^{\,0} $,
rather than that of the initial state, $\sum_k \varepsilon_k^{\,0}\, f_k^{\,0}$. 
In this way, we subtract a trivial fluctuation, related to the random, adiabatic
shift of the single particle energies, and focus on processes where 
particle transitions occur. 

A key observation which allows us to derive closed formulas for the 
work statistics is that for non-interacting fermions, any pure initial state 
is a Slater determinant, which evolves with time 
into  another Slater-determinant, built from time-evolved single particle states  $\varphivec^m(t)$.
As a consequence, quantities such as  instantaneous level occupations as well as 
work, can be expressed in terms of the expansion coefficients $\alpha^m_k(t)$
appearing in the expansion of the time-evolved single particle states 
 in terms of the instantaneous eigenstates, 
\begin{equation}
	\varphivec^m(t)=\sum_{k=1}^N\alpha^m_k(t)\, \etavec_{k,t}\,.
	\label{eq:overlap}
\end{equation}
Once the trajectory of the single particle Hamiltonian ${\cal H}(t)$ is fixed, 
the coefficients $\alpha^m_k(t)$ can be determined numerically by  
solving the single particle Schr\"odinger equation.

As shown in Ref.~\onlinecite{Grabarits1}, the statistical properties of work are most appropriately 
characterized in terms of dimensionless variables such as the dimensionless work, $w=W/\delta \epsilon$, 
the dimensionless time, $\tilde t = t \cdot \delta\epsilon$, 
and the dimensionless velocity, $\tilde v = v/\delta\epsilon/\langle \Delta \lambda \rangle_\mathrm{RM}$, 
with $\delta\epsilon$ denoting the average level spacing, and 
$\langle \Delta \lambda \rangle_\mathrm{RM}$  the dimensionless  separation between Landau-Zener transitions. 
The dimensionless velocity, $\tilde v$, in particular, characterizes the structure of Landau-Zener transitions: 
for $\tilde v\ll1$ no Landau-Zener transitions occur, while for $\tilde v\gg1$ 
the probability of Landau-Zener transitions approaches 1.
At finite temperatures, these three dimensionless  parameters are supplemented 
by the dimensionless temperature, $\tilde T \equiv T/\delta\epsilon$.

\section{Work statistics at $T=0$ temperature}\label{sec:T0}

At $T=0$ temperature,  a diffusively broadened Fermi level is observed~\cite{Grabarits1}
\be
 \langle f_k^{\,T=0}(t)\rangle_\mathrm{RM}  \approx
\Bigl[1-\mathrm{erf}\Bigl(\tilde\varepsilon_k/\sqrt{4\;{\avleft w\avright}_{T=0}}\;\Bigr)\Bigr]/2
\label{eq:f}
\ee
with  $\tilde \varepsilon_k\equiv\left(\varepsilon_k-\varepsilon_\text{F}\right)/\delta\epsilon \approx k-(M+1/2) $ referring to the dimensionless energies measured from  the Fermi energy.  Correspondingly,  the average dimensionless work grows linearly,
\be 
%\label{eq:fandw}
{\avleft w\avright}_{T=0} \approx\widetilde D(\tilde v)\,\tilde t\,,\\
\ee
where  $\widetilde D(\tilde v)$ denotes the dimensionless diffusion coefficient in energy space.
We remark that  $\widetilde D(\tilde v)$  depends only on the dimensionless velocity 
across  Landau--Zener transitions, 
$\tilde v$~\cite{LZ1,LZ2,wilkinson2,wilkinson4,wilkinsondiff1,wilkinson5}, and the underlying  symmetry of the system.
The latter appears through  the anomalous velocity  dependence of $\widetilde D$ for slow quenches, 
$\widetilde D(\tilde v\lesssim1)\sim\tilde v^{1+\beta/2}$. For fast quenches, $\tilde v\gg1$,
 $\widetilde D$ grows quadratically for all ensembles, $\widetilde D(\tilde v\gg1)\sim\tilde v^2$. 
 
 Although energy absorption is a diffusive process in energy space,  work grows super-diffusively
at $T=0$ temperature~\cite{Grabarits1},
 \be
 \label{eq:superdiffusive}
  \delta w^2_{\;T=0} \sim \avleft w\avright^{3/2}.
 \ee
 This is a  manifestation of Fermi statistics combined with
 the diffusive broadening in Eq.~\eqref{eq:f}.

 The central quantity of quantum work statistics is the distribution function, $P(w,\,t)$, 
 characterizing the distribution of work at time $t$. 
  In Ref.~\onlinecite{Grabarits1}, we derived a determinant formula at zero temperature for the characteristic function, 
 \begin{equation}
G(u,\,t)=\int_{-\infty}^\infty\mathrm dw\,e^{-i\,u\,{w}}P(w,\,t)\;,
\end{equation}
 which we expressed at $T=0$  as~\cite{Grabarits1,FeiQuan}
\begin{equation}
\label{eq:detformT0}
G_{T=0}(u, t) =
\left\langle e^{-i\,u\,\sum_{m=1}^M\tilde\varepsilon_m(t)}\mathrm{det}\left[g(u,t)\right]
	\right\rangle_\mathrm{RM}\,.
\end{equation}
Here the $M\times M$ matrix $g(u,t)$ is expressed in terms of the overlaps $\alpha^{m}_k$ in Eq.~\eqref{eq:overlap} as
\begin{equation}
	\left[g(u,t)\right]^{mm^\prime}=\sum_{k=1}^N\left[\alpha^m_k(t)\right]^* e^{i\,u\,\tilde\varepsilon_k(t)}\alpha^{m^\prime}_k(t)
	\label{eq:gu}
\end{equation}
with $m$ and $m^\prime \in \{1,\dots,M\}$.
Eqs.~\eqref{eq:detformT0} and \eqref{eq:gu} generalize Anderson's orthogonality determinant
formula~\cite{Anderson} to work statistics in chaotic Fermi systems. Quantum work statistics can then be analized by 
determining the coefficients $\alpha^m_k(t)$ numerically,  performing the random matrix average 
in Eq.~\eqref{eq:detformT0}, and then performing the inverse Fourier transformation to yield $P(w,\,t)$. 

As discussed in the introduction, $P(w,\,t)$ consists of two distinct parts: an adiabatic piece, $\sim \delta(w)$, and 
a regular part, $P^{\text{reg}}(w,\,t)$.  This structure also  survives  at finite temperatures, 
though $P^\text{ad}(t) $ decays  more rapidly.  

\section{Finite temperature work statistics}\label{sec:T>0}

The $T=0$ temperature concepts of the previous section 
can be carried over to finite temperature with the observation that,  even at finite temperatures, 
the initial state is a thermal average over simple Slater-determinants, evolving into Slater-determinants. 
 We can thus sample the initial states accordingly, while respecting particle number conservation. 

\subsection{Average work}

\begin{figure}[b]
\minipage{0.5\textwidth}
    \includegraphics[width=0.8\textwidth]{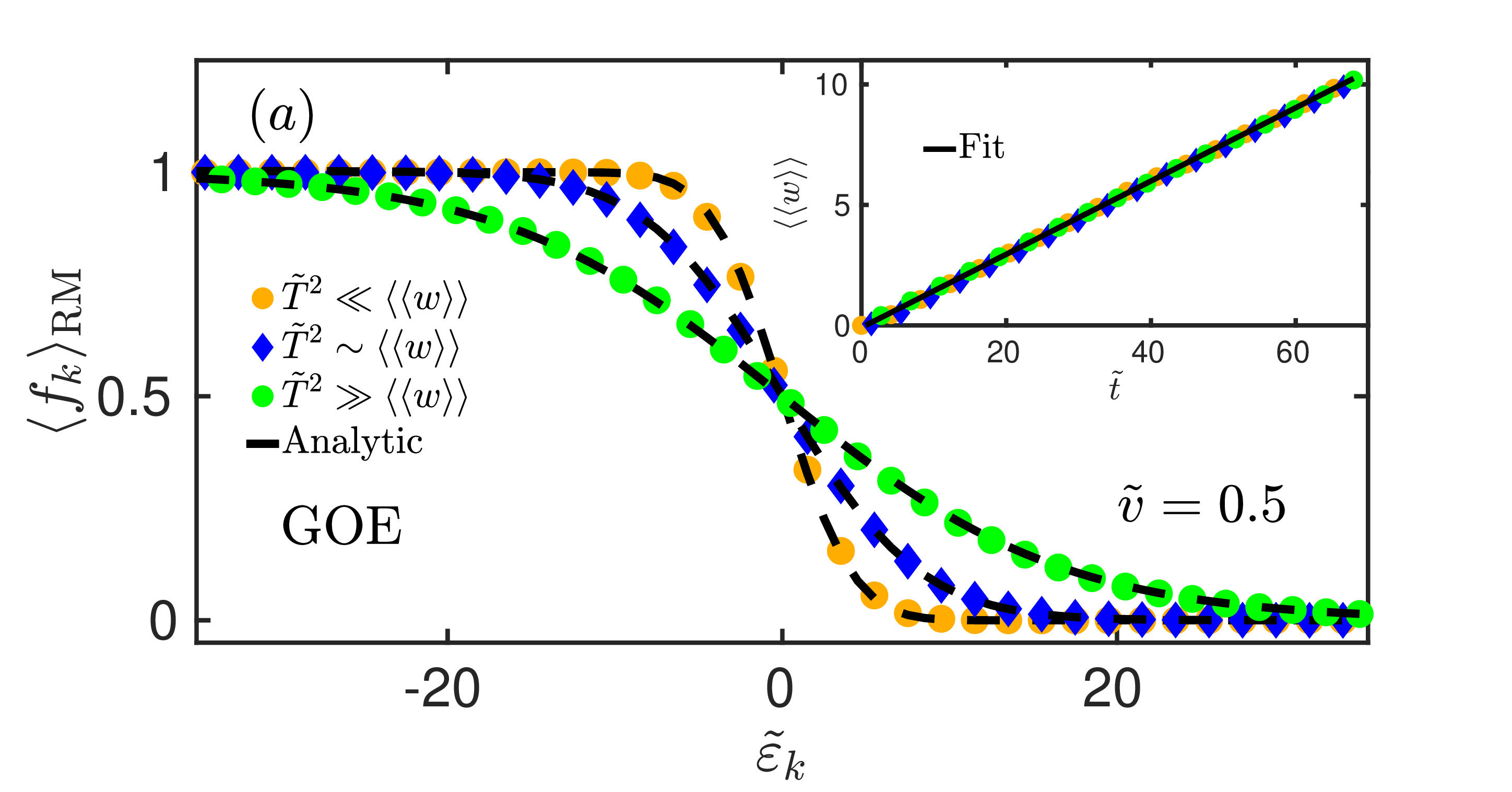}
    \endminipage
  \hfill
\minipage{0.5\textwidth}
    \includegraphics[width=0.8\textwidth]{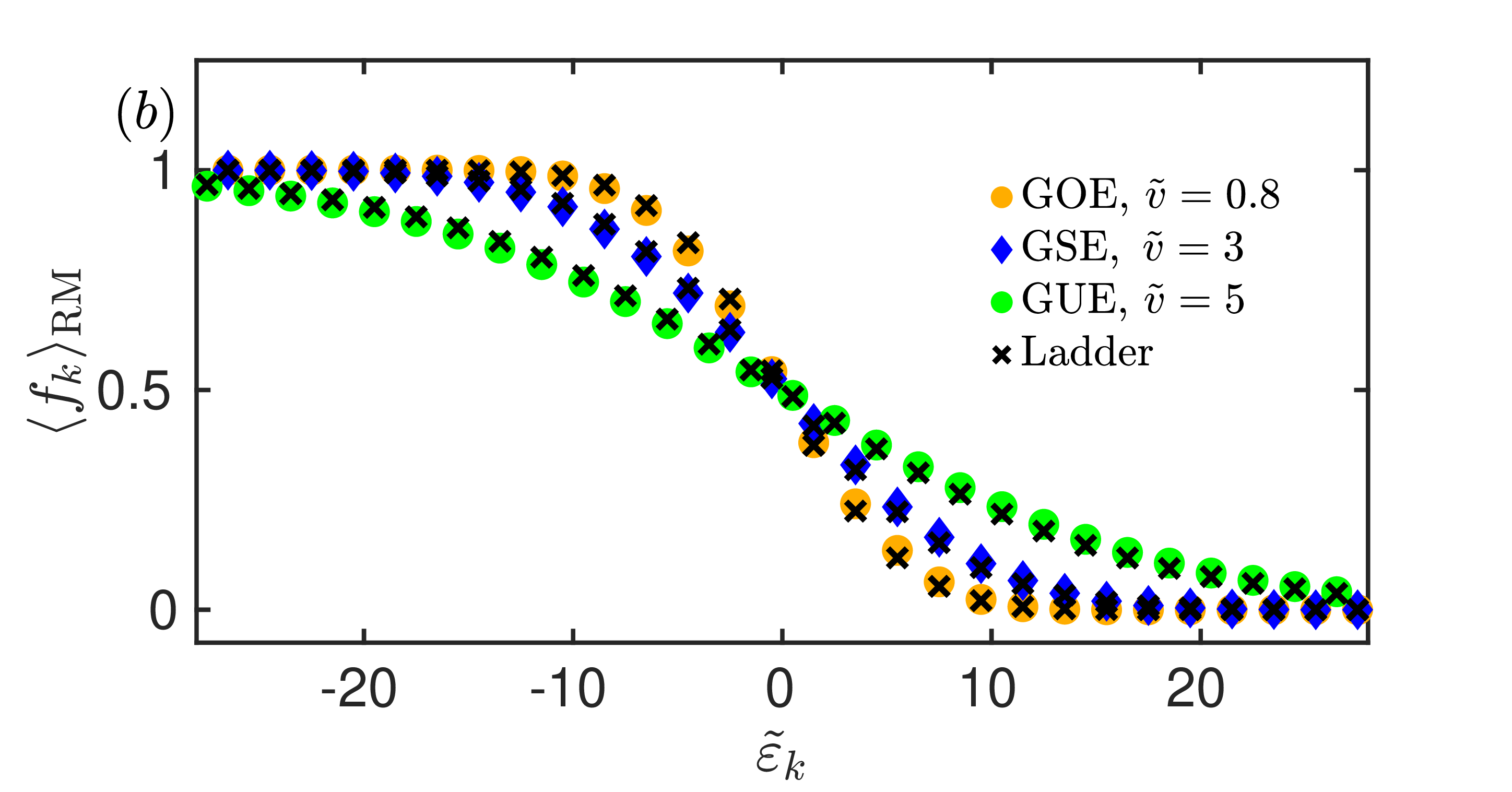}
\caption{Mean occupations $\langle f_k\rangle_\mathrm{RM}$ of instantaneous eigenstates for three different pairs $(\avgw,\tilde T)$, corresponding to large, intermediate, and small work compared to the thermal energy, respectively, and for different velocities $\tilde v$. Quantum mechanical results (symbols), computed from Eq.~\eqref{nteq}, are well captured by the ladder model (black crosses) and  analytical approximation Eq.~\eqref{ncleq}, further simplified to   Eq.~\eqref{eq:lowT} for large, and to Eq.~\eqref{eq:highT}  for small work (black dashed lines). 
(a)  Occupation profiles in the slow quench limit with $\tilde v=0.5$ for the GOE ensemble for 
($\tilde T=0.5$, $\avgw=5$), ($\tilde T=8$, $\avgw=10$)  and ($\tilde T=3,\avgw=10$). Inset:  
 average work as a function of time for the same temperatures $\tilde T$.
{(b)} Occupation profiles for the three random matrix ensembles for different velocities
$(\tilde v=0.8,\tilde T=1,\avgw=10)$ for  GOE, $(\tilde v=3,\tilde T=3,\avgw=15)$ for the GSE, 
 and $(\tilde v=5,\tilde T=8,\avgw=20)$ for  GUE. \label{fig:occupations}}
\endminipage
\end{figure}

To illustrate  emerging complications at $T\ne 0$, let us first consider the level occupations, $f_k(t)$. 
At $T=0$ temperature, these are simply given by the overlap probabilities~\cite{Grabarits1},
$f_k(t) =\sum_{m<M+1}  |\alpha_k^m(t)|^2$,
where the summation is restricted to the occupied levels, i.e. the $M$ lowest lying single particle eigenstates. 
At finite temperatures, this formula is modified as (see Appendix \ref{appfk}
for a detailed derivation),
\begin{equation}\label{nteq}
\begin{split}
	& f_k(t)  =\mathrm{Tr}\left[ \hat n_{k,t} \rho(t)\right]=
	\mathrm{Tr}\left[\hat n^\text{H}_{k,t}\rho_0\right]
	\\
	&=\frac{1}{Z_0(T)}\int_{-\pi}^\pi\frac{\mathrm d\lambda}{2\pi}\,C\argu\lambda\sum_{m}\frac{\lvert\alpha^m_k(t)\rvert^2}{e^{-i
	\lambda+\varepsilon^0_m/T }+1}\;,
	\end{split}
\end{equation}
where $C\argu\lambda=\prod_{l=1}^N\left[e^{-i\lambda f}+e^{i\lambda(1-f)-\varepsilon_l^0/T}\right]$  and $f\equiv M/N$ is the filling factor. Here and throughout the paper we set   $k_B=1$.
Integration over $\lambda$ in Eq.~\eqref{nteq} enforces particle number conservation, while 
 $\hat n^\text{H}_{k,t}= { U}^\dagger(t)\hat n_{k,t}{ U}(t)$ denotes the 
 particle number operator in the $k$-th instantaneous eigenstate in the Heisenberg picture, 
 with ${ U}\argu t=\mathcal T\exp\bigl\{-i\int_0^t\mathrm dt^\prime \hat H\argu{t^\prime}\bigr\}$  the
 many-body  time-evolution operator. 
The trace with the  initial thermal density matrix, $\rho_0=e^{-\hat H\argu0/T}/\,Z_0(T),$
yields  a thermal average over initial  $M$-particle fermionic many-body eigenstates.

Surprisingly, the average work turns out to be independent of the temperature (see inset of Fig.~\ref{fig:occupations}(a)) and assumes
its  $T=0$ value at time $t$, 
\be\label{eq:wt}
\avleft w\avright_{T\ne 0} =  \widetilde D (\tilde v)\; \tilde t
\ee
with $ \widetilde D (\tilde v)$ the dimensionless zero temperature energy diffusion constant, investigated in 
Refs.~\onlinecite{wilkinsondiff1,wilkinson3,wilkinson4,Grabarits1}.

This allows us to replace  time  by the average work, and parametrize the occupation profile  
by $\avleft w\avright$  rather than $\tilde t$  at any temperature.

Fig.~\ref{fig:occupations} shows the computed occupation number profiles for all  three ensembles at low,  high, and intermediate 
 temperatures compared to the injected work, along with 
 the numerical results of the classical symmetric exclusion process simulation and 
 analytical approximations obtained by the `\emph{ladder}' model (discussed in detail in Sec. \ref{sec:SEP}). The behavior of $\langle f_k(t)\rangle_\mathrm{RM}$ depends on the injected work $w$ as compared to the total thermal energy of the 
 fermions, $\sim \tilde T^2$.
 For large works, $\avgw \gg \tilde T^2  $, the broadening is dominantly dynamical, i.e., 
 generated  by electronic  transitions. In this limit,
  the occupation profile 
 converges to the $\tilde T=0$ diffusive profile, Eq.~\eqref{eq:f}.
 % \cite{Grabarits1} $\left\langle\hat n_{k,t}\right\rangle_T\rightarrow\sum_{m=1}^M\lvert\alpha^m_k(t)\rvert^2$, 
 In the small work limit, $\avgw\ll \tilde T^2$, the profile remains a featureless thermal 
 Fermi distribution  $f_k(t) \approx 1 /(e^{(\varepsilon_k^0-\varepsilon_\text{F})/T}+1)$.

\subsection{Finite temperature determinant formula and work statistics}
\label{sec:detform}

We now present analytical expressions for the finite temperature work statistics, 
convenient for numerical evaluation. In particular, we show that similar to the zero temperature case, the work distribution can be determined from the solutions of 
a single particle Schr\"odinger equation, corresponding to the Hamiltonian in Eq.~\eqref{eq:H}. 
Following the convention of the zero temperature case, we introduce the following expression for the PDF,
\begin{equation}
\label{PWT}
	P(w,\,t)=\Big \langle\Big\langle
\delta\left[w-\left(\hat H^\text{H} (t)-\hat H^\text{ad}(t)\right)/\delta\epsilon\right]\Big\rangle \Big \rangle \,.
\end{equation}
Double averaging $\langle\langle\dots\rangle\rangle $ here refers to an average over the initial density operator, 
$\rho_0$, followed by an ensemble average over random matrices.
Here $\hat H^\text{H}(t)=U^\dagger(t)\hat H(t)U(t)$ is the Hamilton operator in the Heisenberg picture, while $\hat H^\text{ad}(t)$ is defined as
\begin{equation}
	\hat H^\text{ad}(t)=\sum_{k=1}^N\varepsilon_k(t)\hat b^\dagger_{k,0}\hat b_{k,0}\,,
\end{equation}
where $\varepsilon_k(t)$ are the instantaneous single particle energy eigenvalues. This operator measures the energy of an adiabatically evolved state. 
Expression \eqref{PWT}  of the work distribution thus
%motivated by the fact that 
corresponds to defining work as the energy of the system relative to the energy of a frozen level occupation profile for each initial state in the thermal average, 
and excludes the contribution of the motion of energy levels.

The characteristic function of the distribution \eqref{PWT} can be expressed in the following form:
\begin{widetext}
\begin{equation}\label{eqdet}
\begin{split}
	G(u,t)&\equiv \int_{-\infty}^\infty\mathrm dw\,e^{-i\,u\,w}P(w,\,t)=
\left\langle\mathrm{Tr}\left[e^{i\,u\,\hat H^\text{H}(t)/\delta\epsilon}e^{-i\,u\,\hat H^\text{ad}(t)/\delta\epsilon} \rho_0\right]\right\rangle_\mathrm{RM}\\
&=\Big \langle
	\frac{1}{Z_0(T)}\sum_n   e^{- i\,u\,E_n(t)/\delta\epsilon} 
	e^{-E^0_n/T}\big\langle\Psi_n(t)\lvert e^{i\,u\,\hat H(t)/\delta\epsilon}\rvert\Psi_n(t)\big\rangle\Big \rangle_
	\mathrm{RM}
	=\Big\langle\frac{1}{Z_0(T)}\sum_n\mathrm{det}\left[g_n(u,t)\right]\Big\rangle_\mathrm{RM}\,.
	\end{split}
\end{equation}
\end{widetext}
Here $|\Psi_n(t)\rangle$ denotes the time-evolved many-body state with some initial occupations $ \{ n_k\}\to n$ 
in the Schr\"odinger picture, where $E_n(t)=\sum_{m=1}^M\varepsilon_{l(m)}(t)$ are the instantaneous many-body energies,
with $\l(m)\in\{1,\dots,N\}$ 
giving the index of the $m$'th occupied  single particle state in the initial many-body state, $|\Psi_n(0)\rangle$.
This is a thermally weighted sum of  $M\times M$ determinants corresponding to the  $M$-particle many-body states. %
The  $M\times M$
overlap matrices $g_n(u,t)$  in Eq.~\eqref{eqdet} are defined similar to the $T=0$ overlap matrix, Eq.~\eqref{eq:gu},
\begin{equation}
\label{eqg}
	\begin{split}
		&\left[g_n(u,t)\right]^{mm^\prime}=\\
		&\sum_{k=1}^N\alpha^{\l (m)}_k(t)e^{i\,u\,\tilde\varepsilon_k(t)}\left[\alpha^{\l(m^
		\prime)}_k(t)\right]^*e^{-i\,u\,\tilde\varepsilon_{\l (m^\prime)}(t)-
		\varepsilon_{\l(m^\prime)}^0/T}
	\end{split}
\end{equation}
with $m$ and $m^\prime\in\{1,\dots,M\}$ running over the particles.
The amplitudes $\alpha$ appearing in Eq.~\eqref{eqg} account for  transitions  between 
these occupied states and the single particle states $k$.
As shown in Appendix \ref{appdet}, 
similar to the occupation in Eq.~\eqref{nteq},
Eq.~\eqref{eqdet} can be written as an integral of a determinant over an auxiliary variable,
\begin{equation}
\label{eqGu}
	G(u,t)=\!\int_{-\pi}^{\pi}\frac{\mathrm d\lambda}{2\pi}\,  
	\Big\langle\frac{ \mathrm{det}\big[ e^{-i\lambda f}+e^{i\lambda(1-f)}\,\mathcal G(u,t)
	\big] }{Z_0(T)}\Big \rangle_{\mathrm{RM}},
\end{equation}
where the matrix elements of the $N\times N$ matrix $\mathcal G(u,t)$ are
\begin{equation}
%\begin{split}
	\mathcal G^{\,\l\, \l^\prime}(u,t)
	=\sum_{k=1}^N\alpha^{\l}_k(t)e^{i\,u\,\tilde
	\varepsilon_k(t)}\left[\alpha^{\l^\prime}_k(t)\right]^*e^{-i\,u\,\tilde\varepsilon_{\l^\prime}
	(t)-\varepsilon_{\l^\prime}^0/T}\,,
%\end{split}
\end{equation}
and $f\equiv M/N$.

\begin{figure}[t!]
\minipage[t]{0.47\textwidth}
    \includegraphics[width=0.9\linewidth,trim={0 0 3cm 1cm},clip]{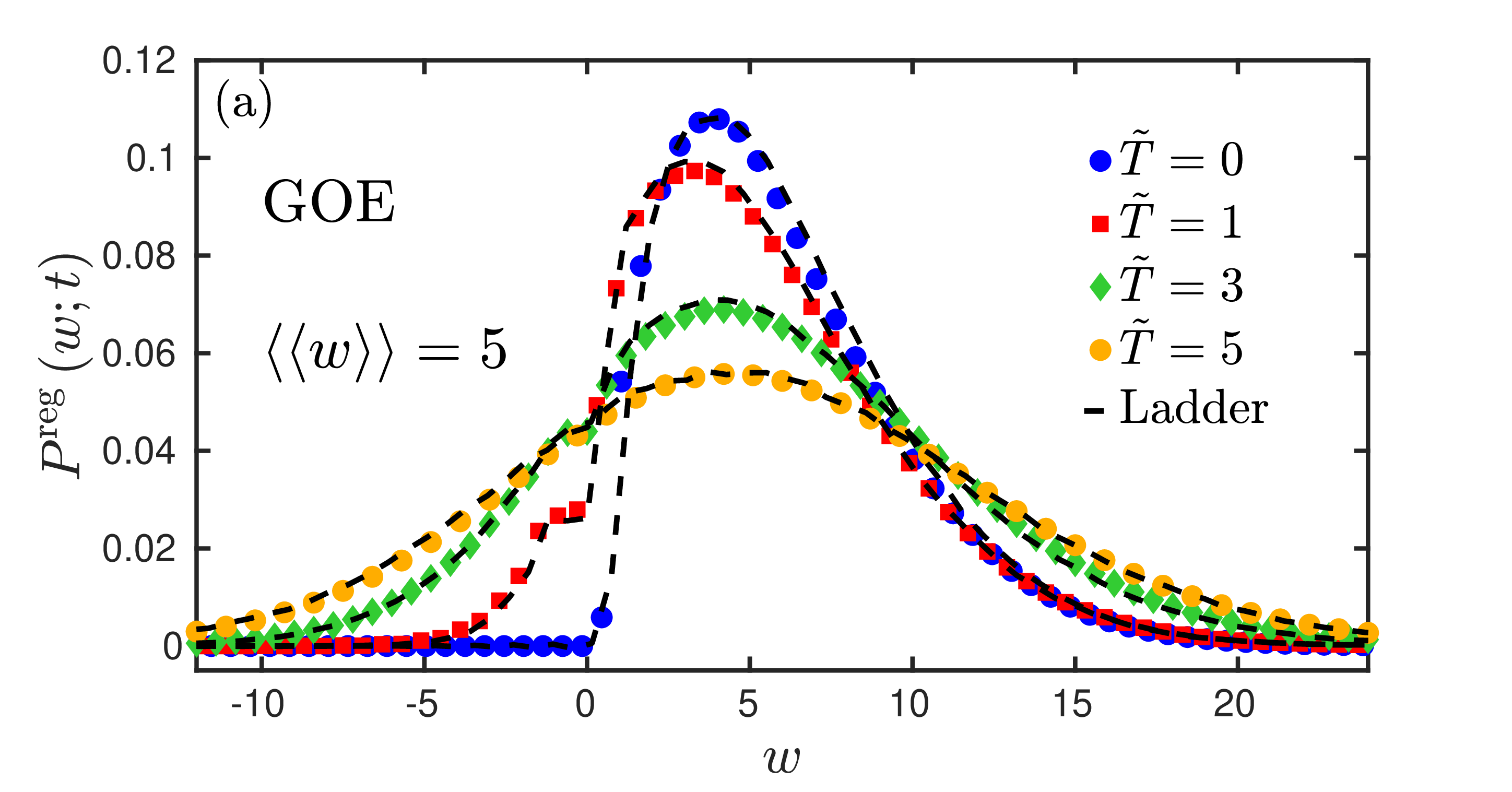}
 \endminipage
  \hfill
\minipage[t]{0.47\textwidth}
    \includegraphics[width=0.9\linewidth,trim={0 0 3cm 0},clip]{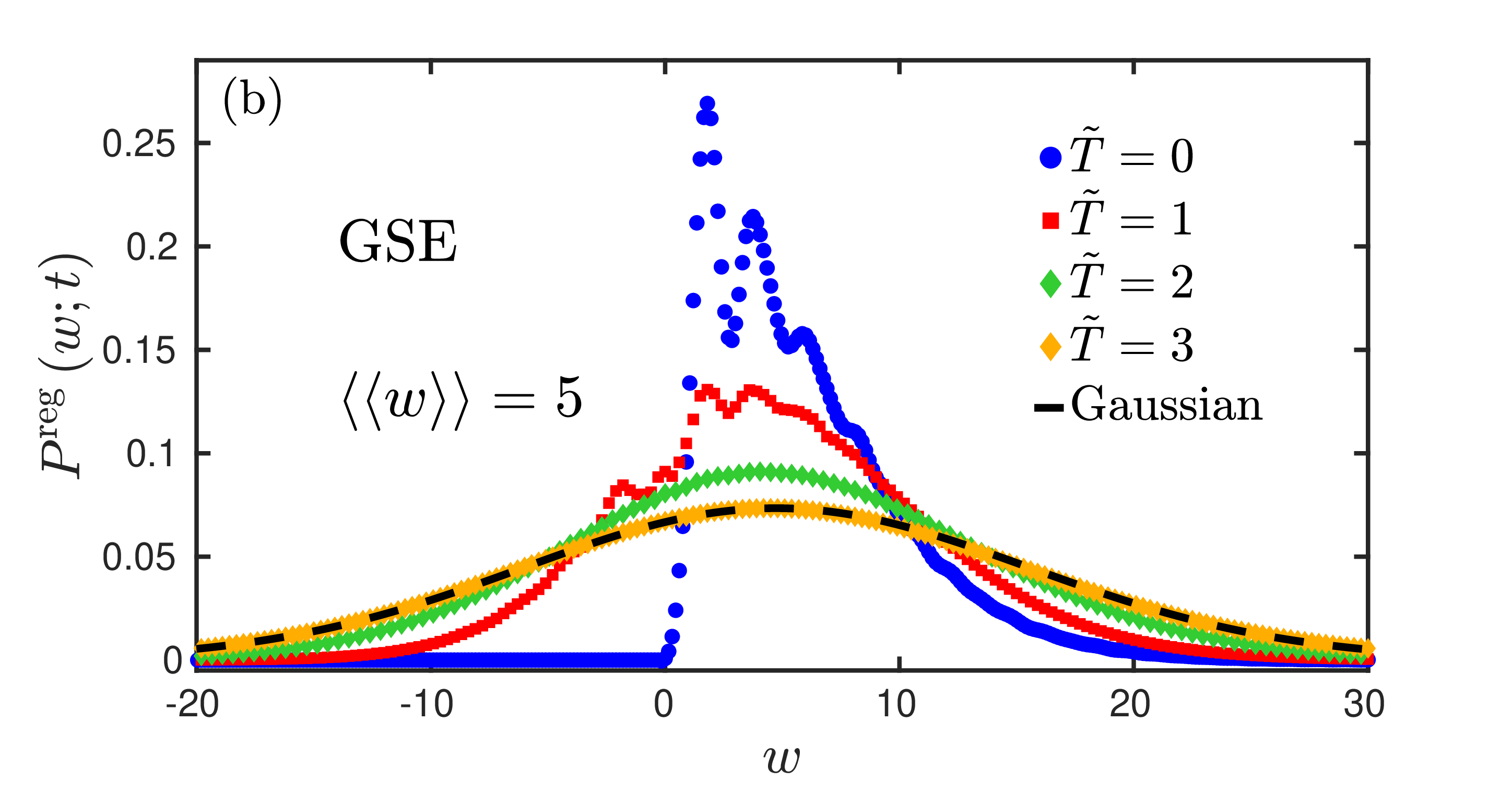}
\endminipage
    \caption{    Regular part of the distribution of quantum work for GOE (up) and GSE (bottom) at various temperatures, for average work $\avgw  =5$, corresponding to velocities  $\tilde v=0.5$ and $0.8$, respectively.
   (a)  Comparison of random matrix simulations  (symbols)
   and the ladder model  (dashed lines) for GOE. Most features are well captured by 
   the classical ladder model.   Fingerprints of level repulsion are present for small 
   positive values of work at low temperatures.
   (b) GSE results.  Level repulsion signatures are enhanced, 
   but disappear fast with increasing  temperature. A fast convergence towards Gaussian statistics is observed
   as $\tilde T^2 \gtrsim \avgw $.     
   }
	\label{fig:GOEPDF}
\end{figure}

As discussed in the Introduction, the work distribution function, obtained as the inverse Fourier transform of the 
characteristic function, $G(u,t)$,   consists of a singular (adiabatic) and a regular part.
We postpone the discussion of the adiabatic part to the next subsection, and focus 
first on the regular part. 

 Typical regular finite temperature work distribution functions for the GOE and GSE ensembles -- obtained by 
a numerical evaluation of Eq.~\eqref{eqGu} followed by  Fourier transformation of the characteristic function -- 
are shown in Fig.~\ref{fig:GOEPDF}~\cite{footnoteB}. 
Unlike at zero temperature, work can now take negative values corresponding to 
transitions from excited states to lower energy states during the quench, taking place 
with a finite probability.

For temperatures and injected work $\avgw$ not too large compared to the mean level spacing, 
the work distribution exhibits fingerprints of level repulsion, 
most clearly visible for GSE, the ensemble with the strongest level repulsion~\cite{Grabarits1}:
the probability distribution displays peaks  at integer multiples of the mean level spacing, 
gradually washed away as $\tilde T$ increases.  
We note that these features are  enhanced in the negative work regime.
 {Nevertheless, in spite of these  more pronounced peaks, the large temperature limit  
 and a smooth distribution is reached faster in  GSE  than in the GOE ensemble
 due to the presence of spin degeneracy, as demonstrated in Fig.~\ref{fig:GOEPDF}.}
In the limit of small works, i.e., for $\tilde T^2\gtrsim \avgw$, 
the work distribution converges towards a Gaussian,
\begin{equation}
	P\argu{w,t}\rightarrow\frac{1}{\textstyle{\sqrt{2\pi\, \delta w^2 }}}\,e^{-\frac{(w-\avleft w\avright )^2}{2 \,\delta w^2  }}\,.
%\phantom{n} \text{ for } \phantom{n} \tilde T\gtrsim \sqrt{\avgw}\,.
\end{equation}
As already stated earlier, the expectation value  $\avgw$ increases linearly as $\avgw = \widetilde D\;\tilde t$, 
with $\widetilde D$ the zero temperature energy diffusion constant. 
This is also reflected in the properties of the characteristic function at $u=0$, where the slope 
remains unaffected by changing   the temperature as shown in  Fig.~\ref{fig:Gu} of Appendix \ref{appdet} 
for several temperatures.

It is an interesting question whether the universal structures in the work distribution,
observed   at zero temperature, survive at finite temperatures.
 The answer is positive:
 for large enough injected work,  work statistics becomes insensitive to most microscopic details, even to 
 the underlying symmetry class and the velocity $\tilde v$, and depends  only on the average work, $\avgw$,
  and the dimensionless initial temperature, $\tilde T$.

\begin{figure}[t]
  \includegraphics[width=0.9\linewidth,trim={0cm 0 0 2cm}]{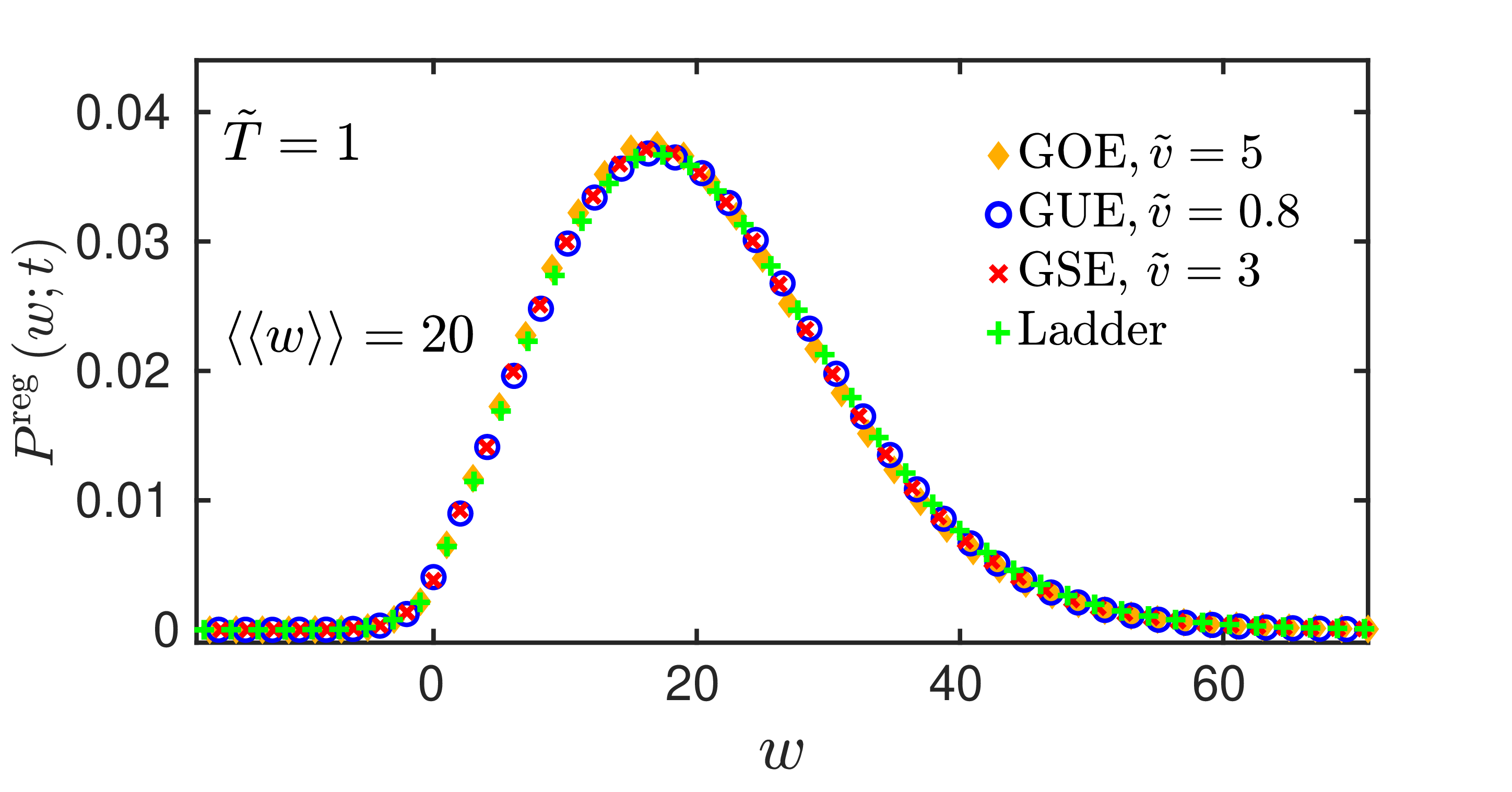}
  \caption{ \label{fig:GOUSE}
  Collapse of work statistics for large average work values and for various velocities within the 
  three major matrix ensembles. The distributions collapse to a single curve, and agree very well 
   with the predictions of the  `ladder' model (green plus signs). Notice the non-Gaussian distribution, 
   characteristic for large works, $\avgw  \gg \tilde T^2$. }
\end{figure}

 This is demonstrated in Fig.~\ref{fig:GOUSE},  where we plot the work distributions for different random matrix
 ensembles and quench velocities at the same temperature and for the same average work. The precise condition for the emergence of universal properties is 
 somewhat different for large and small velocities. For fast quenches, $\tilde v\gg1$, 
when jumps between distant levels occur,
 the injected work must exceed the size of the jumps.
In the  slow quench  limit,  $\tilde v\lesssim 1$, where transitions happen between neighboring energy levels,  the work distribution must be broader than the mean level spacing, while the average work should follow the linear time dependence, Eq.~\eqref{eq:wt}. These conditions imply that for $\tilde v\lesssim 1$, universal features emerge if the average work satisfies 
$\avgw\gtrsim\mathrm{max}\left\{1,\tilde v^{\beta/2-1}\right\}$.

\subsection{Probability of adiabaticity}

Let us now turn to the singular part of the work distribution, corresponding to the probability of adiabatic processes.
As was shown in our previous works~\cite{Grabarits1,Grabarits2}, at $T=0$ the probability of staying in the initial ground state decayed 
in time as a stretched exponential, 
\begin{equation}
	P^\text{ad}_{T=0}(\tilde t)=(8\pi\widetilde D\tilde t)^{1/4}e^{-C\sqrt{\widetilde D\tilde t}},
\end{equation}
with $C\approx1.35$.
In the general finite temperature setting, however, the concept of probability of adiabatic evolution is non-trivial. Following the strategy of the $T=0$ case, we identify it with the weight of the singular part of the work distribution, which, 
based on the properties of the Fourier transformation, can be obtained from the large $u$ asymptotic value of the characteristic function,
\be
\label{eqPad}
P^\text{ad}(t)=\lim_{u\to\infty}G(u,t)\,.
\ee

This corresponds to the sum of probabilities of staying in the initial single particle eigenstates, weighted by the corresponding Boltzmann factors, $e^{-E^0_n/T}/Z_0(T).$ Indeed, Eq.~\eqref{eqPad} can be written as (cf. Appendix \ref{appPad})
\be
\begin{split}
P^\text{ad}(t)&=\sum_n\,\big\langle\frac1{Z_0(T)}\lim_{u\to\infty}\mathrm{det}\left[g_n(t)\right]\big\rangle_\mathrm{RM}\\
&=\sum_n\big\langle\frac{\mathrm{det}\,[\tilde g_n(t)]}{Z_0(T)}\big\rangle_\mathrm{RM}\,,
\end{split}
\ee
where the $u\rightarrow\infty$ limit of the determinants can be expressed as the determinant of the reduced matrix $\tilde g_n(t),$
\be\label{eqgn}
\left[\tilde g_n(t)\right]^{mm^\prime}=\sum_{k=1}^M\alpha^{l(m)}_k\left[\alpha^{l(m^\prime)}_k\right]^*e^{-
\varepsilon^0_{l(m^\prime)}/T}\,,
\ee
yielding exactly the probabilities of staying in the initial $n$th many-body eigenstate, weighted by $e^{-E^0_n/T}$.
Unfortunately, the sum of the above determinants cannot be simplified further, as the matrices $\tilde g_n(t)$ are not simply $M\times M$ blocks of the same $N\times N$ matrix, unlike in the sum of determinants for the characteristic function in Eq.~\eqref{eqdet}. 
We evaluated Eq. \eqref{eqPad} numerically and 
confirmed that it  accounts correctly for the missing weight from the normalization of the work statistics.

\begin{figure}[t]
  \includegraphics[width=0.9\linewidth,trim={0cm 0 0 2cm}]{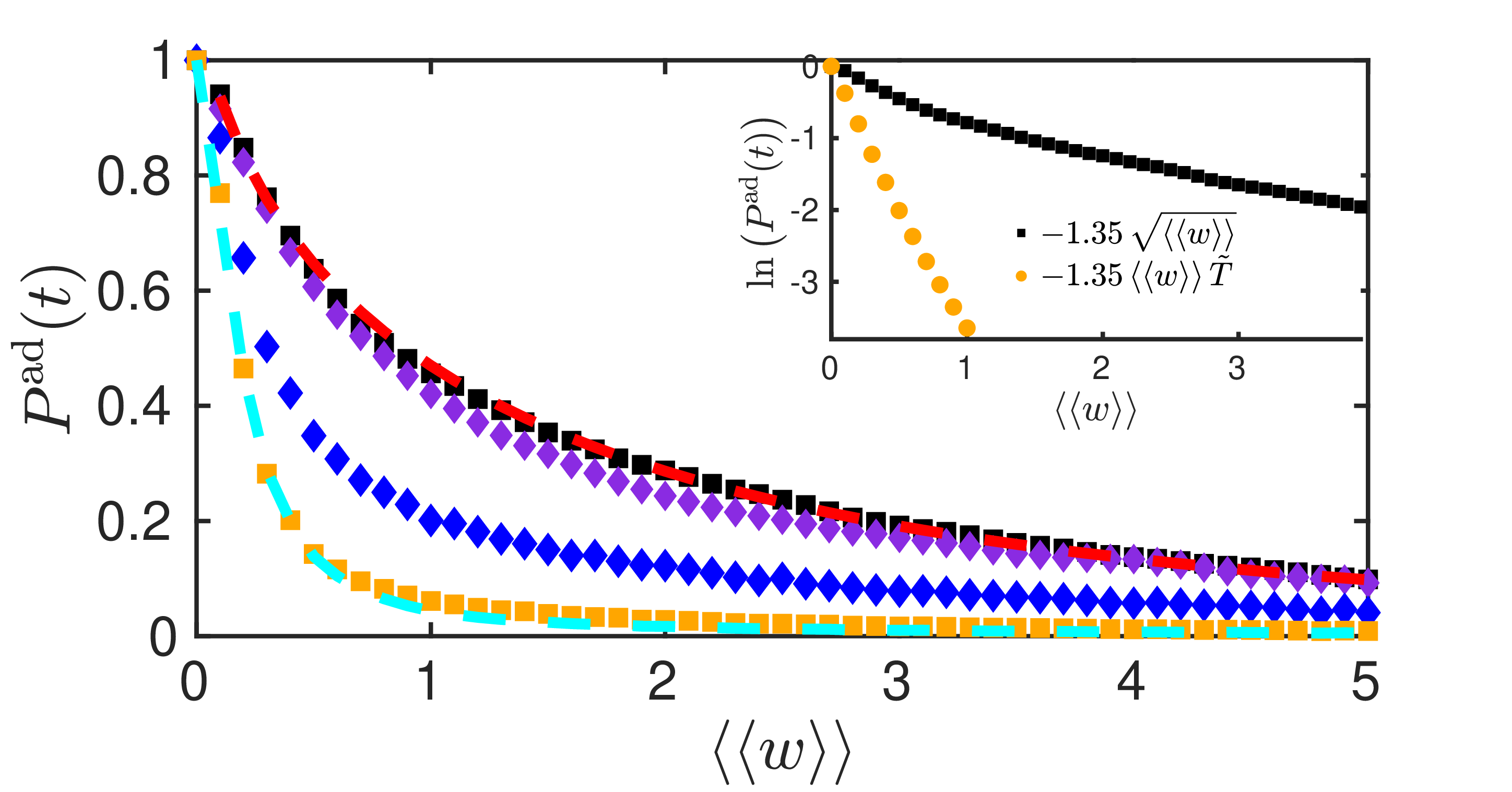}
  \caption{\label{fig:P_ad} Probability of adiabaticity for  GUE  at temperatures $\tilde T=0.5,1.5,3$, for velocity $\tilde v=0.8$ (slow quench limit), and average work $\avgw=5.$  
  At finite $T$, the initial exponential decay crosses over to a stretched exponential 
  for $\avgw> \tilde T^2$. Dashed lines denote analytical exponential and stretched exponential fits.
  Inset:  Logarithmic plot  for $\tilde T=0.5$ and $\tilde T=3$, showing the distinct 
  behaviors at low and high temperatures.}
\end{figure}

Our numerical results for  GUE  are shown in Fig.~\ref{fig:P_ad}. As expected, the stretched exponential behavior is preserved for large work (or low temperatures) $\tilde T^2\ll\avgw$, while we observe a crossover to an exponential decay at small work (high temperatures) $\tilde T^2\gg\avgw$, also highlighted in the inset using logarithmic scales.

For small injected work, $\avgw \ll \tilde T^2$, we find that the probability of adiabaticity decays  exponentially,
\begin{equation}
	P^\text{ad}(\textrm{small}\,W) \propto \exp(-1.35\avgw \,\tilde T) \;,
\end{equation}
where the prefactor remains unchanged  compared to the $T=0$ case up to numerical precision. This result can be understood by noting that in this regime, adiabacity is typically first violated by one of the independent transitions between $\sim T/\delta\epsilon$ partially 
occupied levels close to the Fermi energy, yielding the prefactor in the exponent.

Remarkably, the probability of adiabatic processes remains finite even at finite temperatures. This  allows one 
to perform  mixed state  adiabatic quantum computation. Notice however, that for $\tilde T>1$, one needs to reduce the 
injected work below $1/\tilde T$ by reducing the velocity $\tilde v$ appropriately.

\pagebreak

\subsection{Fluctuations of work and Crooks relation}
\label{sec:workfluct}

In our previous work~\cite{Grabarits1,Grabarits2}, we showed 
 that the fluctuations of work grow superdiffusively at  $T=0$, that is
$\delta w^2 \sim\avgw^{3/2}$.
At finite temperature, we can express the fluctuations in a somewhat more general form as
\begin{widetext}
\begin{equation}\label{Vareq}
	\begin{split}
		\delta w^2&=\left\langle\big\langle\left(H^\text{H}(t)-H^\text{ad}(t)\right)^2\big\rangle-\left\langle H^
		\text{H}(t)-H^\text{ad}(t)\right\rangle^2\right\rangle_\mathrm{RM}\\
	  	&=\sum_{kk^\prime}\left\langle\varepsilon_k(t)\varepsilon_{k^\prime}(t)\left[\left\langle
		\delta\hat n^\text{H}_{k,t}\delta\hat n^\text{H}_{k^\prime,t}\right\rangle+\left\langle\delta\hat n_{k,0}\delta\hat 
		n_{k^\prime,0}\right\rangle-2\left\langle\delta\hat n^\text{H}_{k,t}\delta\hat n_{k^\prime,0}\right\rangle\right]
		\right\rangle_\mathrm{RM}
		\end{split}
	\end{equation}
\end{widetext}
with $\delta\hat n_{k,t}\equiv\hat n_{k,t}-f_k(t)$ and $f_k(t)$ given by Eq.~\eqref{nteq}.
We provide exact expressions for the
correlators appearing here in Appendix~\ref{appavw}, 
in terms of the occupation amplitudes $\alpha^m_k(t)$.

\begin{figure}[b!]
\includegraphics[width=0.45\textwidth]{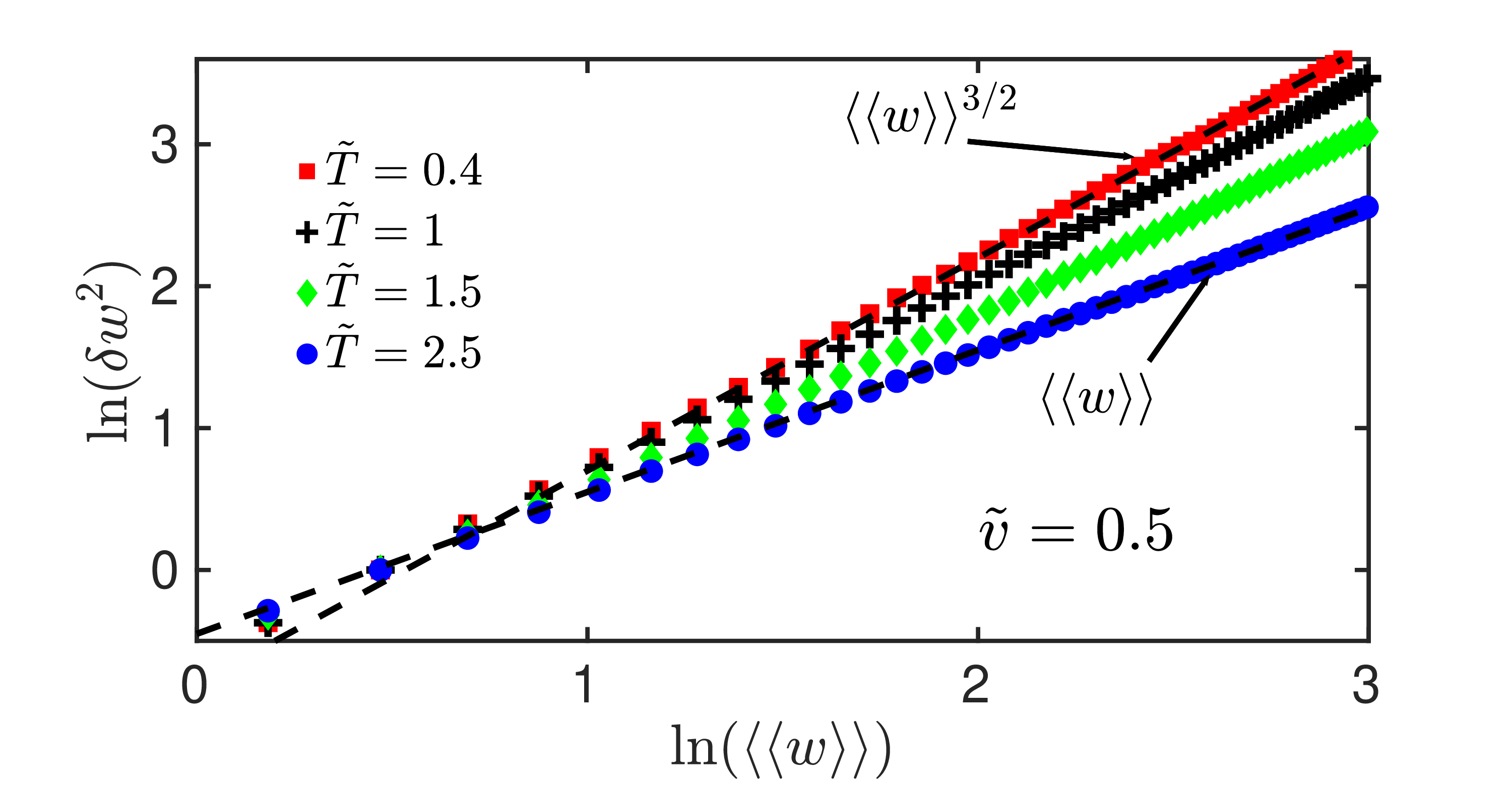}
\caption{Crossover from superdiffusive to diffusive work fluctuations
for $\avgw>1$ in the GOE ensemble.
 For injected work large compared to the thermal energy, fluctuations preserve their superdiffusive behavior as in the $T=0$ case, $\avleft  \delta w^2\avright \sim\avgw^{3/2}$. In contrast, for small average work $1<\avgw  \ll \tilde T^2$, they exhibit  a  diffusive character, $\avleft \delta w^2\avright \sim\avgw$.}
\label{fig:workfluct}
\end{figure}

In the limit of small work, $ 1 <\avgw \ll \tilde T^2$, 
we find a diffusive behavior as  expected, 
\begin{equation}
	 \delta w^2  \sim\avgw \,.  
\end{equation}
For large work, $\avgw \gg \tilde T^2 $, however, 
we recover the superdiffusive characteristics, $ \delta w^2  \sim\avgw^{3/2}$.
Fig.~\ref{fig:workfluct} demonstrates this behavior for  GOE, where for better comparison we introduced the normalized
variance  $\delta\tilde w^2 $ as
\begin{equation}
	 \delta\tilde w^2 =\frac{\delta w^2(\avgw)}{\delta w^2(\avgw=1)}\,,
 \end{equation}
such that $\delta \tilde w^2(\avgw=1) \equiv 1.$ We remark that for large enough injected work, in leading order in 
 $\avgw$, the variance of work coincides
  with that of the total energy of the system, apart from the initial thermal fluctuations of the latter
 (cf. Appendix \ref{appfluct}).

Let us now demonstrate that the numerically computed work distribution  obeys
the celebrated Crooks fluctuation relation~\cite{crooks}.  The Crooks relation relates the work statistics of  
time-reversed processes, $A\to B$ and $ B\to A$, in case of thermal initial states. It states that 
 $P_{A\to B}(W)/P_{B\to A }(-W) = e^{(W-\Delta F)/T}$, where  
$\Delta F$ denotes the equilibrium free energy difference between states $B$ and $A$. 

\begin{figure}[b!]
\includegraphics[width=0.45\textwidth]{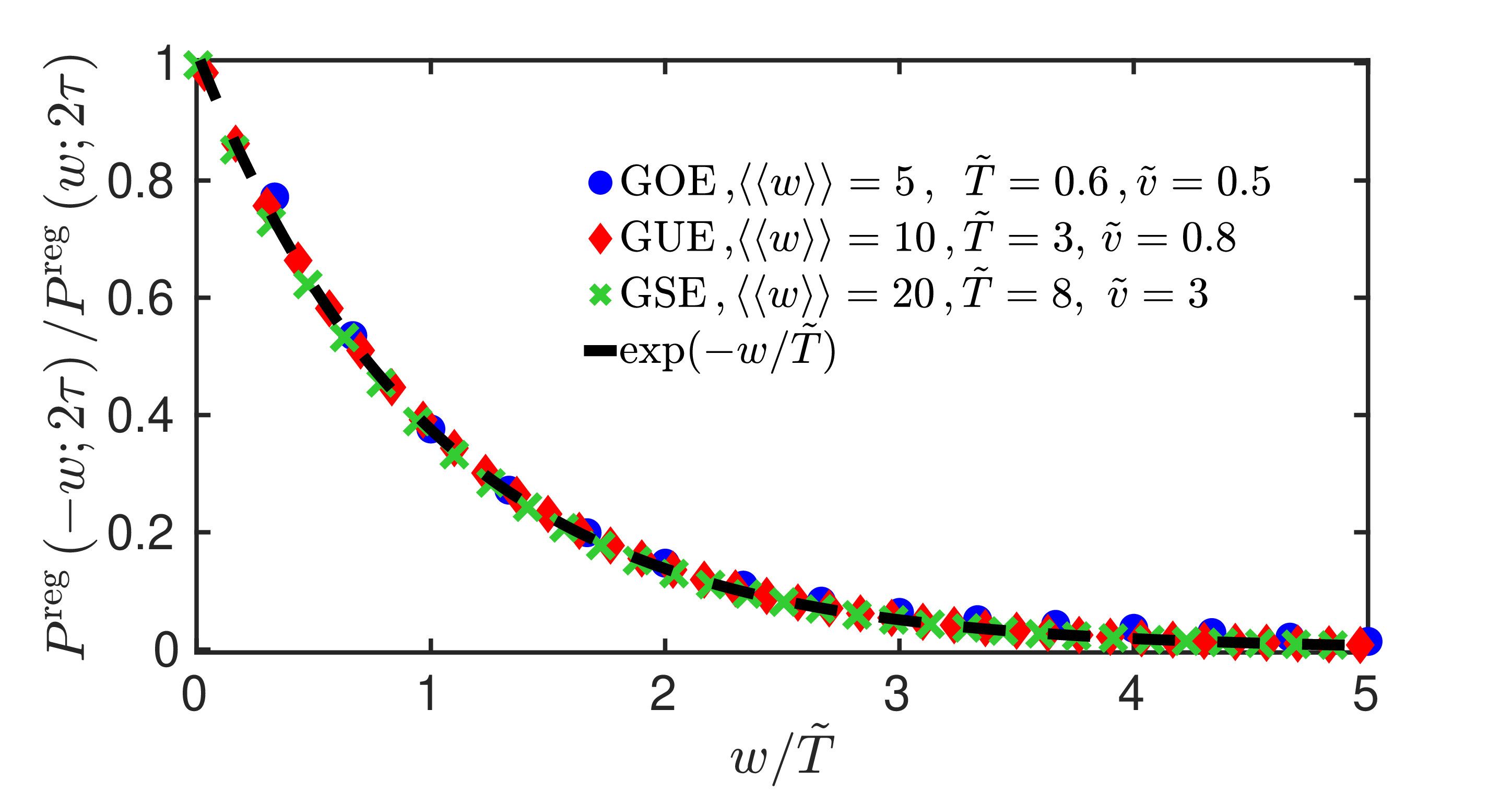}
\caption{Numerical verification of the Crooks fluctuation relation \eqref{eq:crooks} for the three ensembles with different quench times,
velocities, 
and temperatures as indicated in the legend.
The GSE result was computed  using a single random matrix trajectory, 
while the GOE, GUE results were averaged over $\sim 500$ disorder realizations.}
\label{fig:crooks}
\end{figure}

 To check the validity of this relation, we define a time reversal symmetrical 
path in the random matrix ensemble,
\begin{equation}
  \lambda(t) =
\begin{cases}         v\,t   & \text{ for }\;\; 0\leq t\leq\tau\;, \\
      v\,(2\tau-t)   & \text{ for } \;\;\tau\leq t\leq2\tau    \; .     
 \end{cases}
\end{equation} 
In this case, since we return to the same point in parameter space, 
the free energy difference is   $\Delta F=0$, and the Crooks relation 
just states that 
extracting work from the system (performing negative work) is exponentially suppressed with respect to 
performing positive work of the same magnitude as 
\begin{equation}
\label{eq:crooks}
	{P\argu{-w, 2\,\tau }}/{P\argu{w, 2\,\tau}}=e^{-w/\tilde T}\,.
\end{equation}
The two sides of this relation are plotted together in Fig.~\ref{fig:crooks} for all three ensembles at different quench times,
 velocities, and temperatures in each case. Reassuringly, we find that the Crooks relation is satisfied  in all cases within our numerical accuracy. 
 Moreover, we also verified that it remains valid even for a single random matrix realization as it is shown for the GSE ensemble 
 in Fig.~\ref{fig:crooks}. The fact that the Crooks relation is satisfied
is an important verification of the correctness of the rather complicated numerical procedure
used to compute $P (w,\,t)$.

\pagebreak

\section{Simple symmetrical exclusion process and ladder model}

\label{sec:SEP}

\subsection{Finite temperature exclusion process}

The large degree of universality of the work distribution function (cf. Fig.~\ref{fig:GOUSE}) hints at the possibility of an efficient simplified description. This proved to be the case  at zero temperature~\cite{Grabarits1,Grabarits2}, where the work statistics has been captured
 in terms of a \emph{`ladder'} model by a 
simple symmetrical exclusion process~\cite{MarkovChain,LargeIntSystem,StochIntSystem,ASEP}. In this section,
 we show that  the `ladder' model can  successfully  capture   essential features of
  work statistics even at  finite temperatures. 

Within the `ladder' model, we assume infinitely strong level repulsion~\cite{vonDelft_KondoBox_paper,Grabarits2},
and  approximate the spectrum by evenly spaced energy levels. We also 
replace site occupations by \emph{classical} statistical variables taking values $n_{k,t}=0,1.$ Electron 
transfers due to Landau-Zener transitions at  avoided level crossings~\cite{wilkinson1,wilkinson2,wilkinsondiff1}
 are modeled by classical Markovian random jumps between neighboring  energy levels.
At finite temperature, we also need to average over   initial configurations (occupations), 
distributed according to  Boltzmann statistics.
Since the number of configurations is very large  for $N\gtrsim20$ energy levels, we generate  the initial thermal state 
by the Metropolis algorithm, i.e., we move electrons from occupied to empty levels with probability
\begin{equation}
	p=
	\begin{cases}
		&e^{-\Delta E/T}\text{ if }\Delta E>0,\\
		&1\quad\quad\;\;\text{ if }\Delta E<0,
	\end{cases}
\end{equation}
until thermal equilibrium is reached. For each initial state obtained this way, 
%Having performed the first Monte Carlo step for the initial configuration 
we run a second,  stochastic simulation of 
%modelling
 the quantum quench~\cite{Grabarits1},  where electrons diffuse in energy space 
 in a symmetrical exclusion process  due to  driving.  
   Remarkably, as demonstrated in  Fig.~\ref{fig:GOUSE}, for long enough quench times, this 
classical approximation captures work statistics  both in the slow  and fast quench  limits, and
 for all three ensembles.  Quantum statistics and quantum work can thus 
 be described  classically,  analogously to the zero temperature case~\cite{Grabarits2}.

\subsection{Occupations and average work}

In the `ladder' model, we  consider uniformly spaced, fixed energy levels measured from the Fermi energy, 
$\varepsilon_k/\delta\epsilon\approx k-(M+1/2)$. These levels
 do not move in time, though their occupation  changes due to random electron diffusion in energy space, 
induced by the time dependent driving.  The simple structure of this model allows us to obtain analytical 
 results for the energy level occupations as well as for  the mean and variance of  work.

Let us first consider the initial profile. Using $\alpha_k^m(0)=\delta^m_k$ in Eq. \eqref{nteq}, the average occupations are given by the integral expression
\begin{equation}\label{ncl0eq}
	p(k,0) \equiv \langle f_k^0\rangle_\mathrm{RM}\approx\frac{1}{Z_0(T)}\int_{-\pi}^\pi\frac{\mathrm d\lambda}
	{2\pi}\,\frac{C\argu\lambda}{e^{-i\lambda+(k-M-1/2)/\tilde T}+1}\,.
\end{equation}
We can derive an approximate analytical expression for the average occupation numbers at time $t$ by employing a diffusive approximation of the single particle broadenings in Eq. \eqref{nteq}, $\left\langle\lvert\alpha^m_k(t)\rvert^2\right\rangle_\mathrm{RM}\approx{e^{-(k-m)^2/(4\widetilde D\tilde t)}}/{\sqrt{4\pi\widetilde D\tilde t}}$. This approximation simply follows 
from the diffusive motion in energy space, generated by Landau-Zener transitions, as put forward in the 
seminal works of Wilkinson~\cite{wilkinson1,wilkinson2,wilkinsondiff1}.

 Invoking a continuous approximation, whereby $k$ and $m$ are promoted to the continuous variables $\kappa=k-(M+1/2)$ and $\mu=m-(M+1/2)$ and the sum is replaced by integral, we obtain for the occupation profile
\begin{equation}\label{ncleq}
	p(\kappa,t)\approx\int_{-\infty}^\infty\mathrm d\mu\,\frac{e^{-(\kappa-\mu)^2/4\widetilde D\tilde t}}{\sqrt{4\pi
	\widetilde D\tilde t}}p(\mu,0)
\end{equation}
which is just the solution of the diffusion equation with initial profile $p(\kappa,0)$.
 From this equation it follows that the average work grows linearly in time, 
\begin{equation}
	\avgw \approx\int_{-\infty}^\infty\mathrm d\kappa\, \kappa \left[p(\kappa,t)-p(\kappa,0)\right]=\widetilde D\tilde t\,.
\end{equation}
For a derivation, see Appendix \ref{appavw}.
\\

\subsection{Small and large work limits of work fluctuations}\label{subsec:IVC}

We now turn to the discussion of the asymptotic behavior of the occupation profile in the low and high temperature (i.e., large and small work)  
limits. For large works,  $\tilde T^2\ll \avgw$, where the broadening is dominated by driving-induced  level-to-level transitions,
the occupation numbers are given by the zero temperature diffusive profile, with a small correction originating from the initial thermal profile (for the derivation, see Appendix \ref{appasym}),
\begin{equation}
\label{eq:lowT}
	p(\kappa,t)\approx \frac{1}{2}\big(1-\mathrm{erf}\big[ \, {\textstyle \frac{\kappa}{2 \, { \avgw}^{1/2}}}\big]\big)
	+\frac{\pi^2\,\tilde T^2}{6\,\avgw}\frac{\kappa\,e^{-\kappa^2/4\avgw}}{\sqrt{4\pi\avgw}}\,.
\end{equation}
Interestingly, the correction term is twice the Bethe--Sommerfeld correction 
in the absence of particle number conservation.  %to the leading order in $\widetilde T/\sqrt{\avgw}$. 
Eq.~\eqref{eq:lowT} is compared to quantum mechanical results for $\tilde T^2\ll\avgw$ in Fig.~\ref{fig:occupations}a, 
yielding a very accurate approximation.

We can also use this diffusive approximation  to estimate the leading order behavior 
of the work fluctuations for $ \avgw\gg  \max(\tilde T^2, 1)$. This approach keeps track of diffusion-induced  occupation number 
fluctuations, but neglects the randomness of the individual energy levels, as well as  the correlations between 
their occupations,  yielding 
\begin{widetext}
\begin{equation}
\begin{split}
	 \delta w^2(t) 
	%\approx
	%\sum_{k=1}^N \delta\varepsilon^2k^2\left\langle\langle\hat n_{k,t}
	%\rangle\left[1-\langle\hat n_{k,t}\rangle\right]\right\rangle_\mathrm{RM}
	\approx\int_{-\infty}^\infty\mathrm d\kappa\, \frac{\kappa^2}{4}\left(1-\mathrm{erf}^2\left[\kappa/\sqrt{4\,\avgw}   
	\right]\right)+\mathrm o\left(\tilde T^2\avgw^{1/2}\right)
	\approx\frac{5\sqrt\pi}{3}\avgw^{3/2}\;.
	\end{split}
\end{equation}
As shown in Appendix \ref{appfluct}, only the first term of \eqref{Vareq} contributes to the above formula in the leading order of the classical approximation, coinciding with the variance of the total energy.
Our numerics confirm the $\sim \avgw^{3/2}$  power law behavior, but the prefactor is slightly modified by the 
 neglected correlation terms.

In the small work limit, $1<\avgw\ll\tilde T^2$, occupation numbers are approximately given by the Fermi function, plus a correction term accounting for   particle number conservation  and for  quantum quench-induced changes of the
occupation profile (for the derivation, see Appendix~\ref{appasym}), 
%\begin{widetext}
\begin{equation}	
\label{eq:highT}
	p(\kappa,t)\approx\frac{1}{e^{\kappa/\tilde T}+1}+\frac{1}{2\tilde T}\left(\frac{\avgw}
	{2\tilde T}-\frac{1}{4}\right)\, \tanh(\kappa/2\tilde T)\left[1-\tanh^2(\kappa/2\tilde T) \right]
	\,.
\end{equation}
For large enough temperatures,  this expression provides an excellent approximation for the quantum mechanical results and the   ladder model simulations (see Fig.~\ref{fig:occupations}a).

To compute work fluctuations in this limit,  we again  consider only occupation fluctuations of single particle states,
 and neglect their correlations. In Appendix \ref{appfluct}, we show also for the small work case that in the classical approximation only the first term of Eq.~\eqref{Vareq} contributes to the variance of work, thus $\delta^2 w$ reduces to the fluctuations of the total energy. Keeping only the highest order of $\tilde T$, we get
\begin{equation}\label{eq:Var}
	  \delta w^2(t)
	\approx \int_{-\infty}^\infty {\mathrm d}\kappa\,\kappa^2 \, \frac{\avgw}{4\tilde T^2} 
	 \tanh^2(\kappa/2\tilde T)\left[1-\tanh^2(\kappa/2\tilde T) \right]
	 \approx 2.43\,\tilde T\,  \avgw\,.
%	\end{split}
\end{equation}
\end{widetext}

Similar to the large work limit, even though the power law behavior is correct,  correlation terms yield  
 a similar contribution but with a smaller coefficient. The crossover from superdiffusive to diffusive 
 fluctuations is demonstrated in Fig.~\ref{fig:workfluct} both for the quantum mechanical results and the ladder model.

\section{Conclusions}
\label{sec:concl}

In this work, we have extended our earlier, zero temperature results~\cite{Grabarits1,Grabarits2} 
on the  full distribution of quantum work in chaotic 
non-interacting Fermi systems to the case  of  finite temperature quantum quenches. By generalizing 
the determinant formula of Ref.~\onlinecite{Grabarits1},   we have shown that all information is contained 
in the solutions of the single particle Schr\"odinger equation. This enabled us to determine the full work 
distribution function, $P(W,t)$, as well as its moments, and the probability of adiabatic time evolution.

The expectation value of the dimensionless work $w = W/\delta\epsilon$
is found to increase  linearly in time in all three major symmetry classes, 
$$
\langle\langle w\rangle\rangle = \tilde D \, \tilde t\;,
$$
where $\tilde t $ is the dimensionless time, $\tilde t  = t\,\delta\epsilon$, and 
the dimensionless diffusion constant $\tilde D$ depends on the dimensionless speed of the quench
and the symmetry of the underlying Hamiltonian, but is  independent of 
temperature. Let us emphasize that this temperature independence is only valid on time 
scales shorter than one period. For much longer quenches~\cite{wilkinson7}, quantum effects 
such as dynamical localization in energy space~\cite{Kravtsov1,Kravtsov2,Kravtsov3} may become relevant 
and lead to time-dependent, and probably temperature dependent, corrections to the diffusion constant.

Temperature has, however, a dramatic impact on the structure of work distribution. First of all, 
similar  to the effect of large injected work $\avleft W\avright > \delta\epsilon$ at $T=0$ temperature,
a temperature larger than the level spacing, $T>\delta \epsilon$, rapidly washes away `microscopic' features associated with level 
repulsion. 

More importantly, however, at  temperatures $T\gtrsim\delta\epsilon$, the thermal energy of the 
Fermi gas, $E_T\sim T^2/\delta\epsilon$ provides a natural energy scale for quantum work distribution, 
and the properties of quantum work  statistics depend crucially  on the injected energy/work compared to $E_T$.
In the `small work' regime, $\avleft W\avright\lesssim E_T$, quantum statistics  does not play a major role 
in quantum work statistics:  $P(W,t)$ is Gaussian with a very good accuracy, and the  variance of the 
injected work increases linearly with the injected work, $\delta w^2\sim \avgw\sim t$. 
The physical reason for this is that in this `small work' regime, 
 transitions occur in the vicinity of the thermally broadened Fermi energy, and they are statistically independent.
In contrast, in the  `large work' regime, $\langle W\rangle\gtrsim E_T$, work statistics displays 
features similar to those at $T=0$ temperature;   here $P(W,t)$ is generically non-Gaussian due to Fermi statistics, 
and  the variance scales as  $\delta w^2\sim\avgw^{3/2}\sim t^{3/2}$.

The cross-over between  the `small work' and    `large work' regimes is also striking 
in the adiabatic time evolution probability $P_\textrm{ad}(t)$. In the 'small work'  regime, 
$P_\textrm{ad}(t)$ falls off exponentially, 
$$
P_\textrm{ad}(\textrm{small}\,W)\sim e^{- 1.35\avgw \, \tilde T} = e^{-a\, \tilde t}
$$ 
with
$ a\approx1.35\, D \,T / \delta\epsilon^3 = 1.35\tilde D \,\tilde T$. 
For longer times, however, 
where $\langle W\rangle > E_T$, the Pauli principle blocks the transitions, 
and the probability of adiabatic time evolution decreases only as a stretched exponential, 
$$
P_\textrm{ad}(\textrm{large}\,W)\sim e^{- 1.35\,\sqrt{\avgw }}  =   e^{-b\, \tilde t^{1/2}}\,,
$$
where  $b= 1.35\, \sqrt{D / \delta\epsilon} = 1.35\,\sqrt {\tilde D}$.  

These results bare  relevance for adiabatic quantum computation, where the Hamiltonian of a known and simple 
system in some trivial  quantum state  is adiabatically deformed to reach an unknown quantum state
of a more complex system. 
To achieve this,  
one needs to reach a distance $\Delta \lambda = \tilde v\,\tilde t\sim 1 $ in parameter space. 
If the temperature is larger than the level spacing, then the dimensionless work must 
be kept small to have a large adiabatic probability,  $\avgw < 1/\tilde T$, implying a  
small   dimensionless  velocity in parameter space, $\tilde v <1$. In this regime, 
the diffusion constant scales as $\tilde D\sim \tilde v^{1+\beta/2}$, and we obtain the scaling 
$P_\textrm{ad}\sim e^{- \tilde T \, \tilde v ^{\beta/2}\, \Delta \lambda }$. 

Analogously to the $T=0$ temperature case, we find that the above results are well-captured by a 
\emph{classical}  exclusion process model with hardcore particles, moving diffusively in energy 
space, and starting from thermally distributed random initial states.  This 
Markovian approach  reproduces our quantum mechanically obtained work distributions  
and its universal features. These latter imply that work distribution 
in chaotic  fermionic systems is characterized by just two parameters, 
$\avgw$ and $\tilde T$, once $\max(\avgw,\tilde T)\gg  1$,
irrespective of microscopic details,
$$
P(W,t) \to  P(w, \avgw, \tilde T)\;.
$$

Although here we focused on quantum quenches in 
fermionic random matrix ensembles, similar to the $T=0$ temperature case, 
we expect that they apply to most disordered Fermi systems, 
once the appropriate dimensionless parameters are properly identified~\cite{Grabarits2}.
Our results are relevant for quantum computation as well as for quantum thermodynamics, 
and many  of our predictions could be tested experimentally in metallic nanosystems. 

Our considerations  have also many natural extensions; Floquet systems realized by cyclic drivings,
 or  Anderson localization   and interaction effects
 provide exciting  perspectives and future research directions.

\begin{acknowledgments}
We thank P\'eter L\'evay, Anatoli Polkovnikov, and Adolfo del Campo 
for insightful discussions.  This work has
been supported by the National Research Development and Innovation Office (NKFIH) through Grant Nos.
SNN139581 and K138606,  and within the Quantum Information National Laboratory. M.K. acknowledges support by a ``Bolyai J\'anos'' grant of the HAS and by the \'UNKP-21-5 new National Excellence Program of the Ministry for Innovation and Technology from the source of the National Research, Development and Innovation Fund. I.L. acknowledges support from the Gordon and Betty Moore Foundation through Grant GBMF8690 to UCSB.
\end{acknowledgments}

\pagebreak

\onecolumngrid
\appendix
\section{Occupation numbers of instantaneous eigenstates}\label{appfk}
In this appendix we derive the formula Eq.~\eqref{nteq} for the occupation numbers in the instantaneous eigenstates. We prove the general identity at time $t$; from this the special case at $t=0$ follows immediately. First we write out the trace over the $M$-particle initial many-body states with the corresponding Boltzmann weights,
%(for sake of simplicity we drop the $\langle\dots\rangle_\mathrm{RM}$ average and the normalization $Z_0(T)^{-1}$):
%
\begin{equation}\label{eq:A1}
\begin{split}
	f_k(t)&=\frac1{Z_0(T)}\mathrm{Tr}\left[U^\dagger(t)\hat b^\dagger_{k,t}\hat b_{k,t}U(t)e^{-
	\hat H(0)/T}\right]=\frac1{Z_0(T)}\sum_n\big\langle\Psi_n(0)\big\lvert U^\dagger(t)\hat b^\dagger_{k,t}\hat b_{k,t}
	U(t)e^{-\hat H(0)/T}\big\rvert\Psi_n(0)\big\rangle\\
	&=\frac1{Z_0(T)}\sum_ne^{-E^0_n/T}\big\langle\Psi_n(t)\lvert\hat b^\dagger_{k,t}\hat b_{k,t}\rvert\Psi_n(t)
	\big\rangle
	\end{split}
\end{equation}
with $\lvert\Psi_n(t)\rangle=\prod_{m=1}^M\hat c^\dagger_{l(m)}\lvert0\rangle,$ where $l(m)\in\left\{1,\dots, N\right\}$ labels single particle eigenstates, while $m\in\left\{1,\dots,M\right\}$ labels the particles. Now the first task is to evaluate the quantum mechanical average for the arbitrary $n$th many-body state, using the decomposition $\hat b^\dagger_{k,t}=\sum_{m=1}^N\left[\alpha^{m}_k(t)\right]^*\hat c^\dagger_{m,t}$. The result is analogous to the $T=0$ temperature case, but the $M$ lowest levels appearing in the $M$-particle many-body ground state are  now replaced by the general initially occupied states,
\begin{equation*}
	\big\langle\Psi_n(t)\big\lvert\hat b^\dagger_{k,t}\hat b_{k,t}\big\rvert\Psi_n(t)\big\rangle=\sum_{m=1}^M\left\lvert\alpha^{l(m)}_k(t)
	\right\rvert^2.
\end{equation*}
Substituting this into the sum in Eq.~\eqref{eq:A1}, and writing out the summation over the $M$ unequal indices, we transfer the summand to the exponent via the following trick,
\begin{equation*}
\begin{split}
	Z_0(T)f_k(t)=\sum_ne^{-\sum_{m=1}^M\varepsilon^0_{l(m)}/T}\sum_{m=1}^M\left\lvert\alpha^{l(m)}
	_k(t)\right\rvert^2=\frac{\partial}{\partial p}\sum_{l(1)<\dots<l(M)}e^{-\sum_{m=1}^M \varepsilon^0_{l(m)}/T+p\sum_{m=1}
	^M\left\lvert\alpha^{l(m)}_k(t)\right\rvert^2}\Big|_{p=0}\,.
\end{split}
\end{equation*}
We insert the constraint of summing over unequal indices in the form of an integral over the auxiliary variable $\lambda$,% in the following way
\begin{equation*}
	\sum_n\prod_{m=1}^Mh_{l(m)}=\int_{-\pi}^\pi\frac{\mathrm d\lambda}{2\pi}e^{i\lambda M}\,\prod_{l=1}^N\left[1+e^{-i\lambda}h_l\right]=\int_{-\pi}^\pi\frac{\mathrm d\lambda}{2\pi}\,\prod_{l=1}^N\left[e^{i\lambda M/N}+e^{-i\lambda(1-M/N)}h_l\right]\,,
\end{equation*} 
where the two phases ensure that only $M$ number of $h_k$ terms survive the integration. Introducing the filling factor $f\equiv M/N$, performing the differentiation and setting  $p=0$, we obtain the desired formula
\begin{equation}\label{eqB}
\begin{split}
	f_k(t)
	&=\frac{1}{Z_0(T)}\int_{-\pi}^\pi\frac{\mathrm d\lambda}{2\pi}\,\frac{\partial}{\partial p}\prod_{l=1}^N\left[e^{-i
	\lambda f}+e^{i\lambda(1-f)-\varepsilon^0_l/T+p\left\lvert\alpha^l_k(t)\right\rvert^2}\right]\Bigg\rvert_{p=0}\\
	&=\frac{1}{Z_0(T)}\int_{-\pi}^\pi\frac{\mathrm d\lambda}{2\pi}\,\prod_{l=1}^N\left[e^{-i\lambda f}+e^{i\lambda(1-f)-\varepsilon^0_l/T}\right]\sum_{m=1}^N\frac{\left\lvert\alpha^m_k(t)\right\rvert^2}{e^{-i\lambda+
\varepsilon^0_m/T}+1}\,.
\end{split}
\end{equation}
Substituting $\alpha^m_k(0)=\delta^m_k$ if $1\le m\le M$ and 0 otherwise, we obtain the expression for the initial profile:
\begin{equation*}
	f^0_k=\frac{1}{Z_0(T)}\int_{-\pi}^\pi\frac{\mathrm d\lambda}{2\pi}\,\prod_{l=1}^N\left[e^{-i\lambda f}
	+e^{i\lambda(1-f)-\varepsilon^0_l/T}\right]\frac{1}{e^{-i\lambda+\varepsilon^0_k/T}
	+1}\,.
\end{equation*}

\section{Determinant formula for the characteristic function}
\label{appdet}

In this appendix, we derive the determinant formula for the finite temperature characteristic function given in Eq. \eqref{eqdet}. The original expression for the Fourier transform of the work statistics,  Eq.~\eqref{eqGu}, can be understood as a sum of $T=0$ determinant formulae over all the possible $\binom{N}{M}$  initial fermionic many-body sates $\left\{l(1),l(2),\dots,l(M)\right\}$, weighted by the corresponding Boltzmann weights and normalized by the initial partition function, $Z_0\argu T$. This leads to
\begin{equation}
	G\argu{u,t}=\Big\langle
	\frac{1}{Z_0(T)}\sum_n\mathrm{det}\left[g_n(u,t)\right]\Big\rangle_\mathrm{RM}
\end{equation}
with the Boltzmann factors included in the $M\times M$ matrices $g_n\argu{u,t}$ as $\left[g_n(u,t)\right]^{mm^\prime}=\sum_{k=1}^N\alpha^{l(m)}_k(t)e^{i\,u\,\tilde\varepsilon_k(t)}\left[\alpha^{l(m^\prime)}_k(t)\right]^*e^{-i\,u\,\tilde\varepsilon_{l(m^\prime)}(t)-\varepsilon^0_{l(m^\prime)}/T}$.
In this expression, we can extend the summations over all possible $M$-particle states without altering the results, because the determinant vanishes automatically  if two or more rows or columns are the same,
%\begin{widetext}
\begin{equation}\label{eqA1}
\begin{split}
	G(u,t)&\sim\sum_n \mathrm{det}\left[g_n(u,t)\right]=\frac{1}{M!}
	\sum_{\mathcal P}(-1)^{\mathcal P}\sum_{l(1),l(2),\dots,l(M)}g_{l(1)l({\mathcal P_1})}\dots g_{l(M)l(\mathcal P_M)}\\
	&=\sum_{n_1,\dots,n_M}\prod_{i=1}^M
	\frac{(-1)^{n_i+1}((i-1)!)^{n_i}\mathrm{Tr}\left[\mathcal G^i\right]^{n_i}}{(i!)^{n_i}(n_i)!}\delta_{\sum_iin_i-M}\,.
	\end{split}
	\end{equation}
%\end{widetext}
where the matrices $g$ and $\mathcal G$ are defined as
\begin{align*}
	&g_{l(m),l(\mathcal P_m)}=\sum_{k=1}^N\alpha^{l(m)}_k(t)e^{i\,u\,\tilde\varepsilon_k(t)}
	\left[\alpha^{l(\mathcal P_m)}_k(t)\right]^*e^{-i\,u\,\tilde\varepsilon_{l(\mathcal P_m)}-\varepsilon^0_{l(\mathcal 
	P_m)}/T}\,,\\
	&\mathcal G_{ll^\prime}=\sum_{k=1}^N\alpha^l_k(t)e^{i\,u\,\tilde\varepsilon_k(t)}
	\left[\alpha^{l^\prime}_k(t)\right]^*e^{-i\,u\,\tilde\varepsilon_{l^\prime}-\varepsilon^0_{l^\prime}/T}
\end{align*}
with the indices being in the ranges $m\in\left\{1,\dots,M\right\}$ and $l\in\left\{1,\dots,N\right\}.$\\

Here the $\frac{1}{M!}$ factor compensates for the overcounting of states, and the last equation comes from combinatorial reasonings, namely in 
	the sum $\sum_{l(1),\dots,l(M)}\dots$, for all permutations, we can group the product of matrix elements into the form $g_{l(1),l(2)}g_{l(2),l(3)}
	\dots g_{l(i)l(1)}$ which after summation simply yields $\mathrm{Tr}\left[\mathcal G^i\right]$. Suppose that we have $n_i$ such terms that only 
	appear if $\sum_i in_i=M$, as the number of the $g_{n\mathcal P_n}$ elements in a product is fixed. Now in this case the elements in the 
	groups of length $i$ can be permuted among themselves therefore we must compensate with a factor of $\frac{1}{(i!)^{n_i}}$. Then we need to 
	count the permutations of the groups among themselves, which goes with a factor of $\frac{M!}{n_i!}$ and finally, as we have a summation over 
	the permutations each group come up $(i-1)!$ times with the same sign, giving an overall factor of $\frac{((i-1)!)^{n_i}}{(i!)^{n_i}n_i!}=
	\frac{1}{i^{n_i}n_i!}$. As for the sign, in a trace of power $i$, $\mathrm{Tr}\left[\mathcal G^i\right]$ we have $i-1$ exchanges of indices, 
	and altogether we have $n_i$ such terms, resulting in a total sign of $\prod_{i=1}^M\argu{-1}^{n_i\argu{i-1}}=\argu{-1}^M\prod_{i=1}^M\argu{-1}
	^{n_i}\equiv\prod_{i=1}^M\argu{-1}^{n_i+1}$, where we have exploited the constraint $\sum_iin_i=M$.
	
	The trace of the $i$th powers of $\mathcal{G}$ is just the sum of the $i$th powers of its eigenvalues. With some algebra, one can then show that the whole expression of the characteristic function can be written as the sum over $M$ unequal indices ranging from 1 to $N$, $n_1,...,n_m\in{1,...,N}$, where each term is the product of $M$ unequal eigenvalues of the matrix $\mathcal{ G}$. Using this one can rewrite Eq.~\eqref{eqA1} as
\begin{equation}
\begin{split}
	G(u,t)&\sim\sum_{n_1,\dots,n_M}\prod_{i=1}^M\frac{(-1)^{n_i+1}
	\left[\sum_n\mu^i_n\right]^{n_i}}{i^{n_i}(n_i)!}\,
	\delta_{\sum_iin_i-M}=\sum_{n_1<\dots<n_M}\mu_{n_1}\mu_{n_2}\dots\mu_{n_M}\\
	&=\int_{-\pi}^\pi\frac{\mathrm d\lambda}{2\pi}\,\prod_{l=1}^N\left[e^{-i\lambda M/N}
	+e^{i\lambda(1-M/N)}\mu_l\right]=\int_{-\pi}^\pi\frac{\mathrm d\lambda}{2\pi}\,\mathrm{det}\left[e^{-i\lambda f}+e^{i
	\lambda(1-f)}\mathcal G(u,t)\right]\,,
	\end{split}
\end{equation}
 where $\mu_1,\mu_2\dots,\mu_N$ are the eigenvalues of $\mathcal G(u,t)$, and $f=M/N$.
Restoring the prefactor $Z_0(T)^{-1}$ and the $\langle\dots\rangle_\mathrm{RM}$ average, we obtain the desired generalized finite temperature determinant formula in Eq.~\eqref{eqdet}. 

\begin{figure}[t]
\centering
\begin{minipage}{.5\textwidth}
  \centering
  \includegraphics[width=\linewidth,trim={0cm 0cm 4cm 0cm},clip]{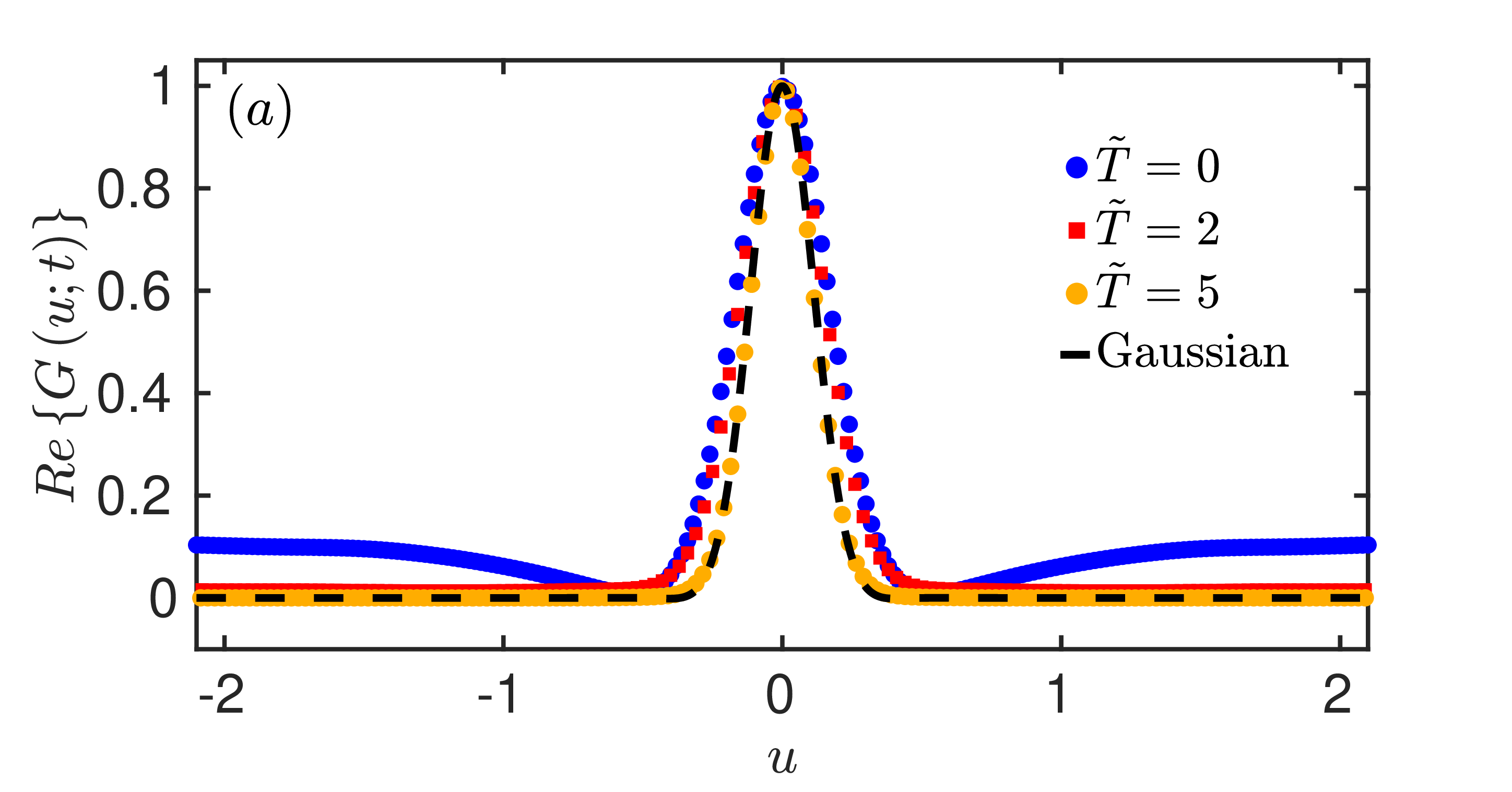}
\end{minipage}%
\begin{minipage}{.5\textwidth}
  \centering
  \includegraphics[width=\linewidth,trim={0cm 0cm 4cm 0cm},clip]{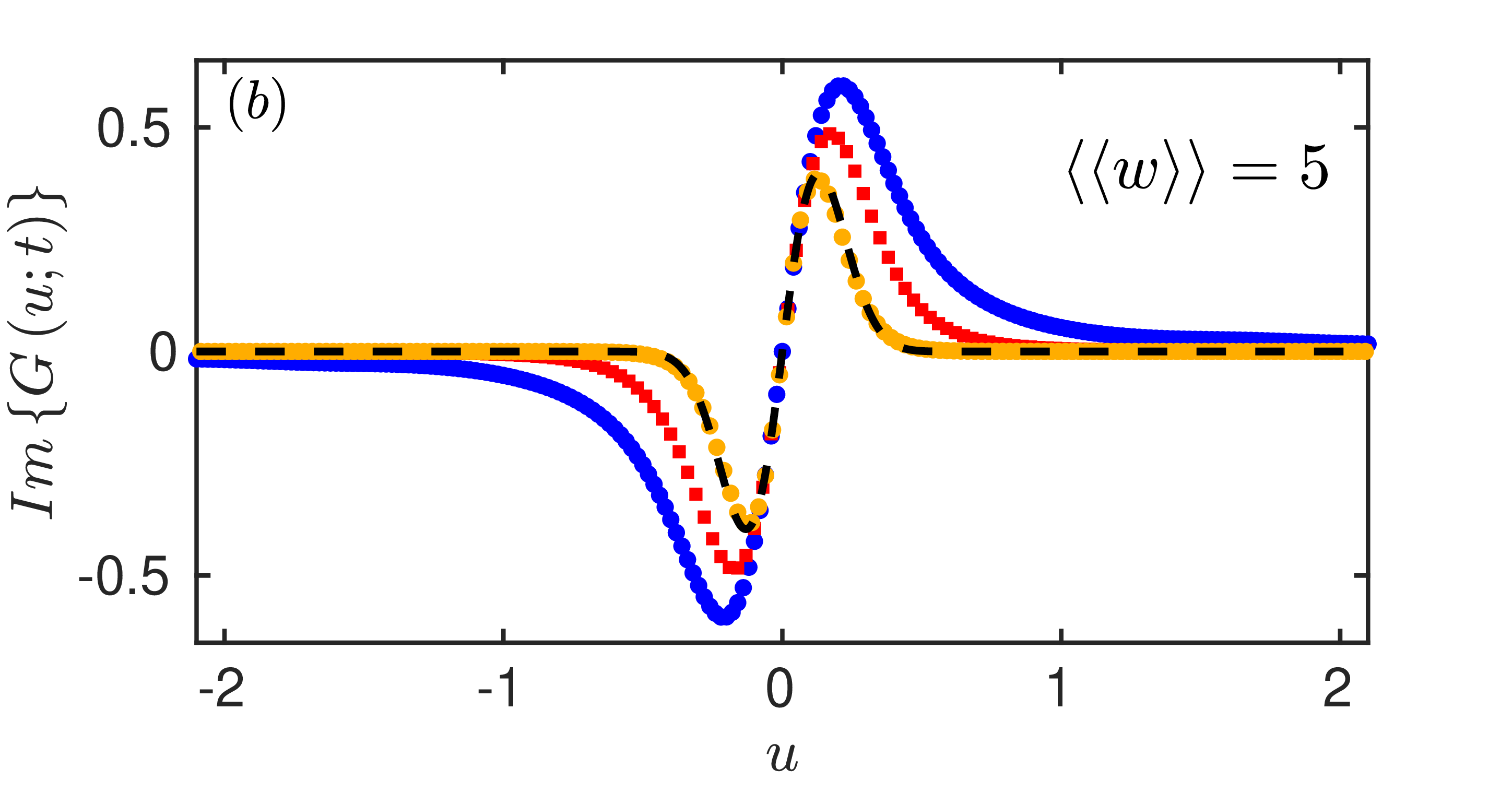}
\end{minipage}
\caption{ Real (a) and imaginary (b) parts of the characteristic function for the GOE ensemble, computed numerically from Eq.~\eqref{eqGu} for dimensionless quench time $\tilde t=34$ and  velocity $\tilde v=0.5$, corresponding to the slow quench limit. The zero temperature average work is $\avgw_{T=0}=5$, and  we evaluate the characteristic function at dimensionless temperatures $\tilde T=0,2,5.$ The slope of the imaginary part at $u=0$ is unaffected by the increase of the temperature, implying that, keeping the quench time fixed, the average work of the system 
%remains intact and so
is independent of the temperature within numerical precision. Upon increasing the temperature, a clear convergence is seen towards the characteristic function of a Gaussian work distribution with mean $\avgw$ and variance $\delta w^2,$ shown in black dashed line.}
\label{fig:Gu}
\end{figure}

The real and imaginary parts of the characteristic function are plotted in Fig.~\ref{fig:Gu} for different temperatures. Note that the slope of the characteristic function at the origin, giving the average work, is independent of the temperature. For $\tilde T^2\gg\avgw,$ the characteristic function converges to the Fourier transform of a Gaussian work distribution.

\section{Probability of adiabaticity}
\label{appPad}

In this appendix we provide the details of computing the probabilities of adiabatic processes starting from a given initial many-body state, and the total adiabatic weight corresponding to the singular part of the work statistics. The probability of adiabatic evolution of the $n$th many body state is given by the absolute value square of the overlap of its initial and time-evolved wave-functions. For sake of brevity, we will drop the time arguments for the energies and $\alpha$ amplitudes. We get
\begin{equation*}
\begin{split}
&\left\lvert\big\langle\Psi_n(0)\big\vert\Psi_n(t)\big\rangle\right\rvert^2=\left\lvert\big\langle0\big\lvert\hat b_{l(M)} \dots\hat b_{l(1)}\hat c^\dagger_{l(1)}\dots\hat c^\dagger_{l(M)}\big\rvert0\big\rangle\right\rvert^2=\left\lvert\sum_{k_1=1}^N\left[\alpha^{l(1)}_{k_1}\right]^*\dots\sum_{k_M=1}^N\left[\alpha^{l(M)}_{k_M}\right]^*\big\langle0\big\lvert\hat b_{l(M)}\dots\hat b_{l(1)}\hat b^\dagger_{k_1}\dots\hat b^\dagger_{k_M}\big\rvert0\big\rangle\right\rvert^2\\
&=
\left\lvert\sum_{\mathcal P}
(-1)^\mathcal{P}\left[\alpha^{l(1)}_{\mathcal P_{l(1)}}\right]^*
\cdots
\left[\alpha^{l(M)}_{\mathcal P_{l(M)}}\right]^*
\right\rvert^2
=\left\lvert\mathrm{det}\left[\alpha_n(t)\right]
\right\rvert^2,
\end{split}
\end{equation*}
where in the last step we used the anticommutation relations of the $\hat b_k,\,\hat b^\dagger_k$ fermionic operators, implying that the non-zero contributions come from indices $\left\{k_1,\dots,k_M\right\}$ being permutations of $\left\{l(1),\dots,l(M)\right\}$. The last expression is the absolute value square of the determinant of the matrix consisting of elements $\left[\alpha_n(t)\right]^m_{m^\prime}=\alpha^{l(m)}_{l(m^\prime)}(t)$. Considering this expression as the product of two determinants, it is equal to the determinant of the product of two matrices, written as $\sum_{k=1}^M\alpha^{l(m)}_{l(k)}\left[\alpha^{l(m^\prime)}_{l(k)}(t)\right]^*$. Now weighting it with the Boltzmann factor, $e^{-E^0_n/T}=e^{-\sum_{m=1}^M\varepsilon^0_{l(m)}/T}$, we get the determinant of the original matrix, with its columns multiplied with Boltzmann weights of the corresponding single particle energies,
\begin{equation}\label{eq:tildegu}
	e^{-E^0_n/T}\left\lvert\big\langle\Psi_n(0)\big\vert\Psi_n(t)\big\rangle\right
	\rvert^2=\mathrm{det}\left[\tilde g_n(t)\right],
\end{equation}
yielding Eq.~\eqref{eqgn}.

We now turn to the $u\rightarrow\infty$ limit of the finite temperature characteristic function, related to the singular part of the work statistics. To this end we extract the $u$ independent part of each  determinant $\mathrm{det}\left[g_n(u,t)\right]$ in the sum determining the characteristic function, Eqs.~\eqref{eqdet} and ~\eqref{eqg}, as this is the only term surviving the limit $u\rightarrow\infty$,
\begin{equation*}
\begin{split}
	&\lim_{u\to\infty}\mathrm{det}\left[g_n(u,t)\right]\\
	&=\lim_{u\to\infty}\sum_{\mathcal P}(-1)^\mathcal{P}
	\sum_{k_1,\dots,k_M=1}^N\alpha^{l(1)}_{k_1}\left[\alpha^{\mathcal P_{l(1)}}_{k_1}\right]^*
	\dots\alpha^{l(1)}_{k_M}\left[\alpha^{\mathcal P_{l(M)}}_{k_M}\right]^*e^{i\,u\left(\tilde
	\varepsilon_{k_1}-\tilde\varepsilon_{\mathcal P_{l(1)}}+\dots+\tilde\varepsilon_{k_M}-
	\tilde\varepsilon_{\mathcal P_{l(M)}}\right)-\left(\varepsilon^0_{\mathcal P_{l(1)}}+\dots+
	\varepsilon^0_{\mathcal P_{l(M)}}\right)/T}\\
	&=\sum_\mathcal{P}(-1)^\mathcal{P}\sum_{k_1,\dots,k_M=1}^M\alpha^{l(1)}_{l(k_1)}
	\left[\alpha^{\mathcal P_{l(1)}}_{l(k_1)}\right]^*\dots\alpha^{l(1)}_{l(k_M)}
	\left[\alpha^{l(\mathcal P_M)}_{l(k_M)}\right]^*e^{-\left(\varepsilon^0_{\mathcal P_{l(1)}}
	+\dots+\varepsilon^0_{\mathcal P_{l(M)}}\right)/T}\equiv\mathrm{det}\left[\tilde g_n(t)\right],
\end{split}
\end{equation*}
where in the last step we used the fact that all the $u$ independent parts are obtained by annullating the energy differences in the exponent of $e^{i\,u\left(\tilde\varepsilon_{k_1}-\tilde\varepsilon_{\mathcal P_{l(1)}}+\dots+\tilde\varepsilon_{k_M}-\tilde\varepsilon_{\mathcal P_{l(M)}}\right)}$ restricting the values of $\left\{k_1,\dots,k_M\right\}$ to the permutations of $\left\{l(1),\dots,l(M)\right\}$. However, we can allow for equal values of $k_1,\dots,k_M$ which give zeros after the summation over the permutations $\mathcal P$, resulting finally in all possible combinations restricted to the set $k_1,\dots,k_M\in\left\{l(1),\dots,l(M)\right\}$. As this holds for all $M$-particle many-body states, we indeed obtain that
\begin{equation*}
	\lim_{u\to\infty} G(u,t)=\sum_n\mathrm{det}\left[\tilde g_n(t)\right].
\end{equation*}
Comparing to Eq.~\eqref{eq:tildegu} confirms that this is equal to the sum of probabilities of adiabatically evolving a particular  initial many-body eigenstate, weighted with the Boltzmann-factors.

\section{Average work}\label{appavw}

In this appendix we show that up to numerical precision, the average work within the classical Markovian approximation is independent of the temperature, $\avgw=\widetilde D\tilde t$. In order to demonstrate this surprising feature, we first realize that the initial profile with the function inside the integral expression, Eq~\eqref{ncl0eq},
\begin{equation*}
	\frac{1}{e^{-i\lambda+\kappa/\tilde T}+1}=\frac{1-\tanh(-i\lambda/2+\kappa/2\tilde T)}{2},
\end{equation*}
is an odd function shifted by a constant.
%by which we have separated the constant part. %and we are left to investigate the 
In the integral in  Eq.~\eqref{ncl0eq}, we exploit that $C\argu\lambda=\prod_{l=1}^N\left[e^{-i\lambda f}+e^{i\lambda(1-f)-\varepsilon_l^0/T}\right]$ is real, thus it is an even function of $\lambda$, $C(-\lambda)=C(\lambda)^*=C(\lambda)$. By switching the integral variable to $\lambda\rightarrow-\lambda$, we obtain
\begin{equation*}
\begin{split}
	&\delta p(-\kappa,0)=\int_{-\pi}^\pi\frac{\mathrm d\lambda}{2\pi}\,C(\lambda)\tanh(-i\lambda/2-\kappa/2\tilde T)
	=\int_{-\pi}^\pi\frac{\mathrm d\lambda}{2\pi}\,C(-\lambda)\tanh(i\lambda/2-\kappa/2\tilde T)\\
	&=-\int_{-\pi}^\pi\frac{\mathrm d\lambda}{2\pi}\,C(\lambda)\tanh(-i\lambda/2+\kappa/2\tilde T)\equiv-\delta p(\kappa,0)\,,
	\end{split}
\end{equation*}
where we denoted by $\delta p(\kappa,t)$ the $\kappa$ dependent part of $p(\kappa,t)$, and exploited that the tangent hyperbolic is an odd function. %and used in the last step the oddity, $\tanh(-x)=-\tanh(k)$.\\

Now we turn to the finite temperature average work as a function of the zero temperature average work, $\avgw_{T=0}=\widetilde D\tilde t$, at time $\tilde t$, where $\widetilde D$ is the diffusion constant. We again replace the summation over the occupation probabilities of the instantaneous eigenstates by a continuous integral, and exploit the fact that $\kappa\, p^{(n\geq1)}(\kappa\rightarrow\pm\infty,0)=0$,
\begin{equation*}
\begin{split}
	\avgw\approx\int_{-\infty}^\infty\mathrm d\mu\,\frac{e^{-\mu^2/4\widetilde D\tilde t}}{\sqrt{4\pi
	\widetilde D\tilde t}}\int_{-\infty}^\infty\mathrm d\kappa\,\kappa\left[p(\kappa-\mu,0)-p(\kappa,0)\right]
	=\int_{-\infty}^\infty\mathrm d\mu\,\frac{e^{-\mu^2/4\widetilde D\tilde t}}{\sqrt{4\pi
	\widetilde D\tilde t}}\int_{-\infty}^\infty\mathrm d\kappa\,\kappa\left[-\mu p^\prime(\kappa,0)+\frac{\mu^2}{2}
	p^{\prime\prime}(\kappa,0)\right].
	\end{split}
\end{equation*}
Here all the odd derivatives drop out, because their product with $\kappa$ is an odd function, while the higher order  even derivatives disappear due to the limiting behaviour of $\kappa p^{(n)}(\kappa,0)$.
Evaluating the $\int\mathrm d\mu\dots$ integral yields $\widetilde D\tilde t$, and we are left with
\begin{equation*}
\begin{split}
	\avgw\approx\widetilde D\tilde t\int_{-\infty}^\infty\mathrm d\kappa\,\kappa p^{\prime\prime}(\kappa,0)=
	\widetilde D\tilde t\left[\kappa p^\prime(\kappa,0)\right]_{-\infty}^\infty
	-\widetilde D\tilde t\int_{-\infty}^\infty\mathrm d\kappa\,p^\prime(\kappa,0)=-\widetilde D\tilde t\left[p(\kappa,0)
	\right]_{-\infty}^\infty=\widetilde D\tilde t
	\end{split}
\end{equation*}
as $p(\kappa\rightarrow-\infty,0)=1$.\\

\section{Small and large work expansion of the occupation numbers}
\label{appasym}

In this appendix we first derive our approximation formula for the occupation profile in the large work limit, $\tilde T^2\ll\avgw$. For sake of simplicity,  we set $M=N/2$, which does not alter our results anyway in the continuum limit. Furthermore, we introduce the reduced partition function $\tilde Z_0(T)=e^{\tilde E^0_\mathrm{GS}/\tilde T}Z_0(T)$ and the corresponding $\tilde C(\lambda)=\prod_{l=1}^{N/2}\left[1+2\cos\lambda\,e^{-(l-1/2)/\tilde T}+e^{-2(l-1/2)/\tilde T}\right]$ in the continuum approximation. We then apply the Bethe--Sommerfeld approach to the modified occupation profile
%namely using the shorthand notation for the $\lambda$ dependent part we apply the expansion now for the "Fermi-like" function 
$\frac{1}{e^{-i\lambda+\mu/\tilde T}+1}$, that is, for a modified chemical potential $\mu\equiv i\lambda\tilde T$,
\begin{equation}\label{eqTlow}
\begin{split}
	&p(\kappa,t)=\frac{1}{\tilde Z_0(T)}\int_{-\pi}^\pi\frac{\mathrm d\lambda}{2\pi}\,\tilde C(\lambda)\int_{-
	\infty}^\infty\mathrm d\mu\,\frac{e^{-\argu{\kappa-\mu}^2/4\widetilde D\tilde t}}{\sqrt{4\pi\widetilde D\tilde t}}
	\frac{1}{e^{-i
	\lambda+\mu/\tilde T}+1}\\
	&\approx\frac{1}{\tilde Z_0(T)}\int_{-\pi}^\pi\frac{\mathrm d\lambda}{2\pi}\,\tilde C(\lambda)\left[\int_{-\infty}
	^{i\lambda\tilde T}\mathrm d\mu\,\frac{e^{-(\kappa-\mu)^2/4\avgw}}{\sqrt{4\pi\avgw}}+\frac{\pi^2\tilde T^2}{6}
	\frac{\argu{\kappa-i
	\lambda\tilde T}}{2\avgw}\frac{e^{-\argu{\kappa-i\lambda
	\tilde T}^2/4\avgw}}{\sqrt{4\pi\avgw}}\right].\\	
\end{split}
\end{equation}
The first term can be calculated by first changing the integration variable as $\mu\rightarrow\kappa+\sqrt{2\avgw}\mu$ and divide the contour into going from $-\infty$ to $-\kappa/\sqrt{2\avgw}$, and then from $-\kappa/\sqrt{2\avgw}$ along the imaginary axis to $\big(-\kappa+i\lambda\tilde T\big)/\sqrt{2\avgw}$,
\begin{equation*}
\begin{split}
	\int_{-\infty}^{-\argu{\kappa-i\lambda\tilde T}/\sqrt{2\avgw}}\mathrm d\mu\,\frac{e^{-\mu^2/2}}
	{\sqrt{2\pi}}=\frac{1-\mathrm{erf}\left[\kappa/\sqrt{4\avgw}\right]}{2}+\int_{-\kappa/\sqrt{2\avgw}}^{-\argu{\kappa-
	i\lambda\tilde T}/\sqrt{2\avgw}}\mathrm d\mu\,\frac{e^{-\mu^2/2}}{\sqrt{2\pi}}.
	\end{split}
\end{equation*}
The complex valued integral appearing here  can be expanded in powers of $i\lambda\tilde T/\sqrt{2\avgw}$ up to order $\sim\tilde T^2/\avgw$,
\begin{equation*}
	\int_{-\kappa/\sqrt{2\avgw}}^{-\argu{\kappa-i\lambda\tilde T}/\sqrt{2\avgw}}\mathrm d\mu\,\frac{e^{-\mu^2/2}}
	{\sqrt{2\pi}}\approx i\lambda\frac{\tilde T}{\sqrt{2\avgw}}\frac{e^{-\kappa^2/4\avgw}}{\sqrt{2\pi}}+\frac{\lambda^2}
	{2}\frac{\tilde T^2}{2\avgw}\frac{\kappa}{\sqrt{2\avgw}}\frac{e^{-\kappa^2/4\avgw}}{\sqrt{2\pi}}.
\end{equation*}
Now the first, purely imaginary part is an odd function of $\lambda$, integrating to zero upon multiplying with the even function $\tilde C\argu\lambda$. The second term is exactly of the same order as the second term in Eq.~\eqref{eqTlow}, li{thus} we can approximate  the $\tilde C\argu\lambda$ and $\tilde Z_0\argu T$ functions by their zeroth order expressions,
\begin{equation*}
	\frac{1}{\tilde Z_0\argu T}\int_{-\pi}^\pi\frac{\mathrm d\lambda}{2\pi}\,\tilde C\argu\lambda\lambda^2\approx\int_{-\pi}^\pi\frac{\mathrm d\lambda}
	{2\pi}\,\lambda^2=\frac{\pi^2}{3}
\end{equation*}
giving in total the contribution $\frac{\pi^2\tilde T^2}{12\avgw}\frac{\kappa e^{-\kappa^2/4\avgw}}{\sqrt{4\pi\avgw}}$ for the first term.
In the second term in Eq.~\eqref{eqTlow}, the terms inside the bracket depend only on the combination $\lambda\tilde T/\sqrt\avgw$. Here we can set $\lambda=0$, as all terms in their expansion around $\lambda=0$ would only contribute in higher orders in the small parameter $\tilde T/\sqrt\avgw$ . This leaves us with the only $\lambda$ dependence coming from the  function $\tilde C\argu\lambda$, resulting in
\begin{equation*}
\begin{split}
	\frac{\pi^2\tilde T^2}{12\avgw}\frac{e^{-\kappa^2/4\avgw}}{\sqrt{4\pi}}
	\int_{-\pi}^\pi\frac{\mathrm d\lambda}{2\pi}\,\frac{\tilde C(\lambda)}{\tilde Z_0(T)}\frac{\kappa-i
	\lambda\tilde T}{\sqrt{\avgw}}e^{\lambda^2\tilde T^2/4\avgw-i\lambda\tilde T\kappa/
	2\avgw}=\frac{\pi^2\tilde T^2}{12\avgw}\frac{\kappa\,e^{-\kappa^2/4\avgw}}
	{\sqrt{4\pi\avgw}}+\mathrm{o}\argu{\tilde T^3\avgw^{-3/2}}.
	\end{split}
\end{equation*}
Together with the previous term, we get a result that is  twice as large as the one without particle number conservation, that is for an initial Fermi distribution,
\begin{equation*}
	p\argu{\kappa,t}\approx\frac{1-\mathrm{erf}\left[\kappa/\sqrt{4\avgw}\right]}{2}+\frac{\pi^2\tilde T^2}{6\avgw}
	\frac{\kappa e^{-\kappa^2/4\langle w\rangle}}{\sqrt{4\pi\avgw}}.
\end{equation*}
Now we turn to the small work, $1<\avgw\ll\tilde T^2$, expansion of the average occupations, with small parameter $\avgw/\tilde T^2\ll1$. Here, for large injected works, $1/\tilde T$ will also be used as small parameter, as series expansions will go only up to leading order in both $1/\tilde T$ and $\avgw/\tilde T^2$. Introducing the new variable $\mu/\tilde T$ and expanding the integral $\int_{-\infty}^\infty\mathrm d\mu\dots$  around $\mu=0$, we get
\begin{equation*}
\begin{split}
	&\int_{-\infty}^\infty\mathrm d\mu\,\frac{e^{-\mu^2\tilde T^2/4\avgw}\tilde T/\sqrt{4\pi\avgw}}{e^{-i\lambda+(\kappa+\mu)/\tilde T}+1}\approx\frac{1}{e^{-i\lambda+\kappa/\tilde T}
	+1}
	\\
	&+\int_{-\infty}^\infty\mathrm d\mu\,\frac{e^{-\mu^2\tilde T^2/4\avgw}\tilde T/\sqrt{4\pi
	\avgw}}{e^{-i\lambda+\kappa/\tilde T}+1}\left[\frac{\mu^2}{(e^{i\lambda-
	\kappa/\tilde T}+1)^2}-\frac{\mu^2/2}{e^{i\lambda-\kappa/\tilde T}+1}\right]\,,
	\end{split}
\end{equation*}
where we exploited that the Gaussian decays rapidly to zero.
Performing the Gaussian integral $\int\mathrm d\mu\dots$, the second term above reads
%resulting together with the $\int\mathrm d\lambda\dots$ integral in
%
\begin{equation*}
	\int_{-\pi}^\pi\frac{\mathrm d\lambda}{2\pi}\,\frac{\tilde C(\lambda)}{\tilde Z_0(T)}\frac{4\avgw}{\tilde T^2(e^{-i
	\lambda+\kappa/\tilde T}+1)}\left[\frac{1/2}{(e^{i\lambda-\kappa/\tilde T}+1)^2}-\frac{1/4}{e^{i\lambda-\kappa/
	\tilde T}+1}\right].
\end{equation*}
We then exponentiate the product defining $\tilde C\argu\lambda$, and replace the sum of the emerging logarithms by a continuous integral, $\tilde C(\lambda)\approx e^{\int_0^\infty\mathrm d\kappa\,\ln\left[1+2\cos(\lambda)e^{-\kappa/\tilde T}+e^{-2\kappa/\tilde T}\right]}$. Now we apply the saddle point approximation, taking into account quadratic corrections of $\tilde C(\lambda)$ and $\tilde Z_0(T)$ around $\lambda=0$, 
$$\tilde C(\lambda)\approx\exp\left\{\int_0^\infty\mathrm d\kappa\,\ln\left[1+e^{-\kappa/\tilde T}\right]-\frac{e^{-\kappa/\tilde T}}{(1+e^{-\kappa/\tilde T})^2}\lambda^2\right\}=e^{\frac{\pi^2\tilde T}{6}-\tilde T\lambda^2/2},$$
$$\tilde Z_0(T)=\int_{-\pi}^\pi\frac{\mathrm d\lambda}{2\pi}\,\tilde C(\lambda)\approx\int_{-\pi}^\pi\frac{\mathrm d\lambda}{2\pi}\,e^{\frac{\pi^2\tilde T}{6}-\tilde T\lambda^2/2}.$$ 
After some algebraic manipulation, the correction terms read as
\begin{equation*}
\begin{split}
	&\frac{2\avgw e^{\frac{\pi^2\tilde T}{6}}}{\tilde T^2\tilde Z_0(T)}\int_{-\pi}^\pi
	\frac{\mathrm d\lambda}{2\pi}\,e^{-\tilde T\lambda^2/2}\frac{1-\tanh^2(-i
	\lambda/2+\kappa/2\tilde T)}{4}\left[\frac{1}{e^{i\lambda-\kappa/\tilde T}+1}-\frac{1}{2}\right]=
	\frac{\avgw}{4\tilde T^2\tilde Z_0(T)}\int_{-\pi}^
	\pi\frac{\mathrm d\lambda}{2\pi}\,e^{-\tilde T\lambda^2/2}\\
	&\times\left[1-\tanh^2(-i\lambda/2+\kappa/2\tilde T)\right]\tanh(-i\lambda/2+
	\kappa/2\tilde T)=\frac{\avgw}{4\tilde T^2}\left[1-\tanh^2(
	\kappa/2\tilde T)\right]\tanh(\kappa/2\tilde T)+\mathrm o\argu{\avgw/\tilde T^{5/2}},
	\end{split}
\end{equation*}
where we could neglect the $\lambda$ dependence in the $\tanh(\dots)$ function, as expanding it around $\lambda=0$ yields higher order correction terms in the small parameter $1/\tilde T$ for large enough values of $\avleft w\avright$. Following a similar procedure with the remaining initial profile integral, we get
\begin{equation*}
	\begin{split}
	&\frac{1}{\tilde Z_0(T)}\int_{-\pi}^\pi\frac{\mathrm d\lambda}{2\pi}\,\tilde C(\lambda)\frac{1}{e^{-i
	\lambda+\kappa/\tilde T}+1}\approx\frac{1}{e^{\kappa/\tilde T}+1}-\frac{1}{\tilde Z_0(T)}e^{\frac{\pi^2}{6}
	\tilde T}\int_{-\pi}^\pi\frac{\mathrm d\lambda}{2\pi}\,e^{-\tilde T\lambda^2/2}\frac{1}{e^{\kappa/\tilde T}+1}
	\left[\frac{\lambda^2}{(e^{-\kappa/\tilde T}+1)^2}-\frac{\lambda^2/2}{e^{-\kappa/\tilde T}+1}\right]\\
	&=\frac{1}{e^{\kappa/\tilde T}+1}-\frac{1}{8\tilde T}\left[1-\tanh^2(\kappa/2\tilde T)\right]
	\tanh(\kappa/2\tilde T)+\mathrm o\argu{1/\tilde T^{3/2}}\,.
	\end{split}
\end{equation*}
In the last steps, having evaluated all Gaussian integrals, the given approximation is also applied to $\tilde Z_0(T)$. The final result, together with the previous terms, yields the desired expansion valid for works small compared to the temperature, $1<\avgw\ll\tilde T^2$,
\begin{equation}
	p(\kappa,t)\approx\frac{1}{e^{\kappa/\tilde T}+1}+\left(\frac{\avgw}{4\tilde T^2}-\frac{1}{8\tilde T}\right)\left[1-
	\tanh^2(\kappa/2\tilde T)\right]\tanh(\kappa/2\tilde T).
\end{equation}

\section{Correlation terms and variance of work}\label{appfluct}
This appendix is devoted to the details of the calculations leading to the analytical expressions obtained for the correlators $\langle\delta\hat n^\text{H}_{k,t}\delta\hat n^\text{H}_{k^\prime,t}\rangle$ and $\langle\hat n^\text{H}_{k,t}\hat n_{k^\prime,0}\rangle$. We also discuss the leading order behavior of the classical approximate formula for the variance of work.\\
Starting with the correlator terms appearing in the variance of work, we generalize our zero temperature result for the two-point correlators of occupation number fluctuations expressed as a function of the expansion amplitudes $\alpha$. We show that the expression below provides a formula convenient for numerical implementation,
\begin{equation*}
\begin{split}
	&\langle\delta\hat n^\text{H}_{k,t}\delta\hat n^\text{H}_{k^\prime,t}\rangle=\langle\hat n^\text{H}_{k,t}\hat 
	n^\text{H}_{k^\prime,t}\rangle-f_k(t)f_{k^\prime}(t)=\mathrm{Tr}
	\left[\left(\hat b^\dagger_{k,t}\right)^\text{H}\hat b^\text{H}_{k,t}\left(\hat b^\dagger_{k^\prime,t}\right)^
	\text{H}\hat b^\text{H}_{k^
	\prime,t}\,\rho_0\right]-\mathrm{Tr}\left[\left(\hat b^\dagger_{k,t}\right)^\text{H}\hat b^\text{H}_{k,t}
	\rho_0\right]
	\mathrm{Tr}\left[\left(\hat b^\dagger_{k^\prime,t}\right)^\text{H}\hat b^\text{H}_{k^\prime,t}\rho_0\right]\\
	&=-\frac{1}{Z_0(T)}\int_{-\pi}^\pi\frac{\mathrm d\lambda}{2\pi}\,\prod_{l=1}^N\left[e^{-i\lambda f}
	+e^{i\lambda(1-
	f)-\varepsilon^0_l/T}\right]\left\lvert\sum_{m=1}^N\frac{\left[\alpha^m_k(t)\right]^*\alpha^m_{k^\prime}(t)}
	{e^{-i\lambda+\varepsilon^0_m/T}+1}\right\rvert^2\left(1-\delta_{kk^\prime}\right)+f_k(t)(1-f_k(t)\delta_{kk^
	\prime}\,.
	\end{split}
\end{equation*}
Here again $\hat n_{k,t}=\hat b^\dagger_{k,t}\hat b_{k,t}$ denotes the particle number operator of the $k$th state in the instantaneous basis, with $\hat n^\text{H}_{k,t}$ being in Heisenberg picture. The two Kronecker deltas imply that the only non-trivial part is $k\neq k^\prime$, while for $k=k^\prime$ we naturally have $\big\langle\left(\delta\hat n^\text{H}_{k,t}\right)^2\big\rangle=f_k(t)(1-f_k(t))$ for fermions. For  $k\neq k^\prime$, we can write expressions similar to the $T=0$ case, but with different initial many-body states weighted with the corresponding Boltzmann probabilities, 
\begin{equation*}
		\langle\delta\hat n^\text{H}_{k,t}\delta\hat n^\text{H}_{k^\prime,t}\rangle=-\frac{1}{Z_0(T)}
		\sum_ne^{-\sum_{m=1}^M\varepsilon^0_{l(m)}/T}\left\lvert\sum_{m=1}^M\alpha^{l(m)}_k(t)\alpha^{l(m)}_{k^\prime}(t)
		\right\rvert^2\,.
\end{equation*}
We can rewrite this again by differentiating a sum of exponentials over the fermionic $M$-particle many-body states twice,
\begin{equation*}
\begin{split}
	&-\frac{1}{Z_0(T)}\frac{\partial}{\partial p}\frac{\partial}{\partial p^\prime}\sum_{l(1)<\dots<l(M)}e^{-\sum_{m=1}^M
	\left(\varepsilon^0_{l(m)}/T+p\left[\alpha^{l(m)}_k(t)\right]^*\alpha^{l(m)}_{k^\prime}(t)+p^\prime\alpha^{l(m)}_k(t)
	\left[\alpha^{l(m)}_{k^\prime}(t)\right]^*\right)}\Bigg\rvert_{p,p^\prime=0}\\
	&=-\frac{1}{Z_0(T)}\int_{-\pi}^\pi\frac{\mathrm d\lambda}{2\pi}\,\frac{\partial}{\partial p}\frac{\partial}{\partial p^
	\prime}\prod_{l=1}^N\left[e^{-
	i\lambda f}+e^{i\lambda(1-f)-\varepsilon^0_l/T+p\left[\alpha^l_k(t)\right]^*\alpha^l_{k^\prime}(t)+p^\prime
	\alpha^l_k(t)\left[\alpha^l_{k^\prime}(t)\right]^*}\right]\Bigg\rvert_{p,p^\prime=0}\\
	&=-\frac{1}{Z_0(T)}\int_{-\pi}^\pi\frac{\mathrm d\lambda}{2\pi}\,\prod_{l=1}^N\left[e^{-i\lambda f}+e^{i\lambda/
	2(1-f)-\varepsilon^0_l/T}\right]\left\lvert\sum_{m=1}^N\frac{\left[\alpha^m_k(t)\right]^*
	\alpha^m_{k^\prime}(t)}{e^{-i\lambda+\varepsilon^0_m/T}+1}\right\rvert^2\,,
	\end{split}
\end{equation*}
Indeed, this gives the right formula for the nontrivial case of $k\neq k^\prime$.

Turning to the non-equal time correlator $\langle\hat n^\text{H}_{k,t}\hat n_{k^\prime,0}\rangle$,
first we calculate it for a given initial state. Analogously to the $T=0$ case, we get
\begin{equation*}
	\left\langle\Psi_n(0)\left\lvert\hat n^\text{H}_{k,t}\hat n_{k^\prime,0}\right\rvert\Psi_n(0)
	\right\rangle=\left\langle\Psi_n(t)\left\lvert\hat n_{k,t}\right\rvert\Psi_n(t)\right\rangle n_{k^\prime}=\sum_{m=1}^M
	\left\lvert\alpha^{l(m)}_k(t)\right\rvert^2n_{k^\prime},
\end{equation*}
with $n_{k^\prime}=0,1$ constraining the summation to many-body states $\left\lvert\Psi_n(0)\right\rangle$ with  the $k^\prime$th single particle state occupied, i.e., $l(m)=k^\prime$ for some $m\in\left\{1,\dots,M\right\}$. Now inserting this into the expression of the occupation number correlator, we can write
\begin{equation*}
	\sum_ne^{-E^0_n/T}\,n_{k^\prime}\sum_{m=1}^M\left\lvert\alpha^{l(m)}_k(t)\right\rvert^2=\frac{\partial}{\partial p}
	\sum_nn_{k^\prime}\,e^{-\sum_{m=1}^M\varepsilon^0_{l(m)}/T+p\left\lvert\alpha^{l(m)}_k(t)\right\rvert^2}\Big\vert_{p=0}.
\end{equation*}
This expression can again be transformed into an integral over the auxiliary variable $\lambda$, enforcing particle number conservation, but with the additional constraint  $l(m)=k^\prime$ for some $m\in\left\{1,2,\dots,M\right\}$. This is accomplished by bringing out the term $e^{i\lambda(1-f)-\varepsilon^0_{k^\prime}/T}$ in front of the product, then multiply both the nominator and denominator with the missing term, $e^{-i\lambda f}+e^{i\lambda(1-f)-\varepsilon^0_{k^\prime}/T}$, yielding
\begin{equation*}
	\begin{split}
		&\langle\hat n^\text{H}_{k,t}\hat n_{k^\prime,0}\rangle=\frac{1}{Z_0(T)}\frac{\partial}{\partial p}\int_{-\pi/2}
		^{\pi/2}\frac{\mathrm d\lambda}{2\pi}\,\prod_{l=1}^N\left[e^{-i\lambda f}+e^{i\lambda(1-f)-\varepsilon^0_l/T
		+p\left\lvert
		\alpha^l_k(t)\right\rvert^2}\right]\frac{1}{e^{-i\lambda+\varepsilon^0_{k^\prime}/T}+1}\Big\vert_{p=0}\\
		&=\frac{1}{Z_0(T)}\int_{-\pi/2}^{\pi/2}\frac{\mathrm d\lambda}{2\pi}\,\prod_{l=1}^N\left[e^{-i\lambda f}+e^{i\lambda(1-f)-\varepsilon^0_l/T}\right]\frac{1}{e^{-i\lambda+\varepsilon^0_{k^\prime}/T}
		+1}\sum_{m=1}^N\frac{\left\lvert
		\alpha^m_k(t)\right\rvert^2}{e^{-i\lambda+\varepsilon^0_m/T}+1}.
	\end{split}
\end{equation*}
This expression is convenient for studying  the limits $\tilde T^2\gg\avgw$ and $\tilde T^2\ll\avgw$ in the classical approximation.

Turning to the limiting behavior of the variance of work, the key observation is that in the continuum limit, the non-equal time correlator terms, $\langle\delta\hat n^\text{H}_{k,t}\delta\hat n_{k^\prime,0}\rangle$, disappear in leading order for both the large and small work limiting cases, $\avgw\gg\tilde T^2$ and $\avgw\ll\tilde T^2$, respectively. Note that, similarly to the zero temperature case~\cite{Grabarits1,Grabarits2}, it is enough to focus on the $k=k^\prime$ case, giving the main contributions to the fluctuations of work. Starting with the large work limit, $\tilde T^2\ll\avgw$, we perform the same expansion of the integral expression of the diagonal part of the correlator, $\langle\hat n^\text{H}_{k,t}\hat n_{k,0}\rangle$,  as in the previous appendix in the classical limit. Moreover, we approximate the integrand by the product of the $T=0$ diffusion term and the factor$\frac{1}{e^{-i\lambda+\kappa^\prime/\tilde T}+1}$, with the latter implementing the constraint,
\begin{equation*}
\begin{split}
		&\int_{-\pi/2}^{\pi/2}\frac{\mathrm d\lambda}{2\pi}\,\frac{C(\lambda)}{Z_0(T)}\frac{1}{e^{-i
		\lambda+\varepsilon^0_{k^\prime}/T}+1}\sum_{m=1}^N\frac{\left\lvert\alpha^m_k(t)\right\rvert^2}
		{e^{-i
		\lambda+\varepsilon^0_m/T}+1}\approx\int_{-\pi/2}^{\pi/2}\frac{\mathrm d\lambda}{2\pi}\,
		\frac{1}{e^{-i\lambda+\kappa^\prime/\tilde T}+1}\frac{1-\mathrm{erf}\left[\kappa/\sqrt{4\avgw}\right]}{2}\\
		&+\mathrm o\big(\tilde T^2/\avgw
		\big)=\frac{1-\mathrm{erf}\left[\kappa/\sqrt{4\avgw}\right]}{2}\Theta(-\kappa^\prime)+\mathrm o\big(\tilde T^2/\avgw
		\big)+\mathrm o\big(e^{-\lvert \kappa^\prime\rvert/\tilde T}\big).
		\end{split}
\end{equation*}
Here, in the last step, we expanded $\frac{1}{e^{-i\lambda+\kappa^\prime/\tilde T}+1}$ around $\tilde T=0$, up to exponential accuracy for temperatures that are small compared to the the average works. Up to leading order, this expression is precisely the product of the leading order approximations of $f_k(t)\approx\frac{1-\mathrm{erf}\left[\kappa/\sqrt{4\avgw}\right]}{2}+\mathrm o\big(\tilde T^2/\avgw\big)$ and $f^0_{k^\prime}\approx\Theta(-\kappa^\prime)+\mathrm o\big(e^{-\lvert\kappa^\prime\rvert/\tilde T}\big)$, leading to $\langle\delta\hat n^\text{H}_{k,t}\delta\hat n_{k^\prime,0}\rangle=\langle\hat n^\text{H}_{k,t}\hat n_{k^\prime,0}\rangle-f_k(t)f^0_{k^\prime}\sim\mathrm o\big(e^{-\lvert\kappa^\prime\rvert/\tilde T}\big)$, not contributing in highest order to the $\sim\avgw^{3/2}$ work variances.

Turning to the small work limit, $1<\avgw\ll\tilde T^2$, it is enough to consider the terms depending on the average work, as these are the only terms contributing to the variance of work. For both the off-diagonal and diagonal terms, the expansion of the single particle states' occupation number  in the classical approximation yields for the average work dependent part up to leading order
\begin{equation*}
	f_k(t)f^0_{k^\prime}\approx\frac{\avleft w\avright}{4\tilde T^2}\left[1-
	\tanh^2(\kappa/2\tilde T)\right]\tanh(\kappa/2\tilde T)\frac{1}{e^{\kappa^\prime/\tilde T}+1}.\\
\end{equation*}
In the case of the non-equal time correlator terms, we  expand the integral along the lines of the calculation of the occupation numbers. Finally, in the last step, we insert the constraint via the term $\frac{1}{e^{-i\lambda+\kappa^\prime/\tilde T}+1}$, arriving at
\begin{equation*}
\begin{split}
	&\int_{-\pi/2}^{\pi/2}\frac{\mathrm d\lambda}{2\pi}\,\frac{C(\lambda)}{Z_0(T)}\frac{1}{e^{-i\lambda+\varepsilon^0_{k^\prime}/T}+1}\sum_{m=1}^N\frac{\left\lvert\alpha^m_k(t)\right\rvert^2}{e^{-i\lambda+
	\varepsilon^0_k/T}+1}\\
	\approx\frac{\avgw}{4\tilde T^2\tilde Z_0(T)}&\int_{-\pi}^\pi\frac{\mathrm d\lambda}{2\pi}\,e^{\frac{\pi^2}{6}\tilde T-
	\frac{\lambda^2}{2}\tilde T}\frac{1}{e^{-i\lambda+\kappa^\prime/\tilde T}+1}\left[1-\tanh^2(-i\lambda/2+	\kappa/
	2\tilde T)\right]\tanh(-i\lambda/2+\kappa/2\tilde T).
	\end{split}
\end{equation*}
From here we can repeat the reasoning used for the calculation of the occupation numbers, and take $\lambda=0$ everywhere except  the first exponential, as we are already at order $\sim\frac{\avgw}{\tilde T^2}$, and all expansions of $\lambda$ would just yield higher orders $\sim\mathrm o\left(\avgw/\tilde T^{5/2}\right)$. In the last step we also expand the partition function up to the same order, $\tilde Z_0(T)\approx\int\frac{\mathrm d\lambda}{2\pi}\,e^{\tilde T\pi^2/6-\tilde T\lambda^2/6}$, leaving us with the final result
\begin{equation*}
	\langle\hat n^\text{H}_{k,t}\hat n_{k^\prime,0}\rangle\approx\frac{\avgw}{4\tilde T^2}\left[1-
	\tanh^2(\kappa/2\tilde T)\right]\tanh(\kappa/2\tilde T)\frac{1}{e^{\kappa^\prime/\tilde T}+1}.
\end{equation*}
This cancels the product of averages  in leading order, $\langle\delta\hat n^\text{H}_{k,t}\delta\hat n_{k^\prime,0}\rangle\sim \mathrm o\left(\avgw/\tilde T^{5/2}\right)$, giving no contribution to the largest order of variances, $\sim\avgw\tilde T$.

 The variance is then calculated from the lowest order term depending on the average work, coming from 
\begin{equation*}
	f_k(t)\left(1-f_k(t)\right)\approx p(\kappa,t)\left(1-p(\kappa,t)\right)=\frac{\avgw}{4\tilde T^2}\tanh^2(\kappa/
	2\tilde T)\left[1-\tanh^2(\kappa/2\tilde T)\right]+\mathrm o\left(\avgw/\tilde T^{5/2}\right),
\end{equation*}
matching exactly the integrand in \eqref{eq:Var}.\\
 Now one can observe that, apart from the initial thermal energy fluctuations, $\delta E^2_0(T)\sim\tilde T^3$, the variance of the total energy of the system grows according to the same power law behavior as the variance of work in the classical picture and within numerical precision. To verify this statement, let us write down the variance of energy in terms of the single particle states' correlators (again for brevity dropping the random matrix ensemble average, $\langle\dots\rangle_\mathrm{RM}$),
\begin{equation*}
	\left\langle\left(\delta H^\text{H}(t)\right)^2\right\rangle=\left\langle\left(H^
	\text{H}(t)\right)^2\right\rangle-\left\langle H^\text{H}(t)\right\rangle^2=\sum_{kk^
	\prime}\varepsilon_k(t)\varepsilon_{k^\prime}(t)\left\langle
	\delta\hat n^\text{H}_{k,t}\delta\hat n^\text{H}_{k^\prime,t}\right\rangle.
\end{equation*}
This variance consists of an average work dependent part and a purely temperature dependent one, now not cancelling out at $t=0$. The large $\avgw$ part originates from the approximating expression
\begin{equation*}
	\sum_k\tilde\varepsilon^2_k(t)f_k(t)(1-f_k(t))\approx\int_{-
	\infty}^\infty\mathrm d\kappa\,\kappa^2 p(\kappa,t)(1-p(\kappa,t)),
\end{equation*}
giving the same result as for $\delta w^2$ in highest order. Here we again focused only on occupation fluctuations, and neglected the randomness of energy levels  and the correlation terms. As for thermal fluctuations, we indeed obtain via similar approximations
\begin{equation*}
	\left\langle\left(\delta H^\text{H}(0)\right)^2\right\rangle\approx\delta\epsilon^2\int_{-\infty}^\infty\mathrm d
	\kappa\,\kappa^2 p(\kappa,0)\left(1-p(\kappa,0)\right)\sim \tilde T^3
\end{equation*}
in leading order, as the approximate expression for $p(k,0)$, Eq.~\eqref{ncl0eq}, contains the variable $\kappa$ multiplied with the inverse temperature, $\kappa/\tilde T$.


\begin{references}

\bibitem{Hanggi}P. Talkner, E. Lutz, and P. H\"anggi, Phys. Rev. E \textbf{75}, 050102(R) (2007).
\bibitem{workreview}M. Campisi, P. H\"anggi, and P. Talkner, Rev. Mod. Phys. \textbf{83}, 771 (2011). 

\bibitem{jarz}C. Jarzynski, Phys. Rev. Lett. \textbf{78}, 2690 (1997).
\bibitem{crooks}G. E. Crooks, Phys. Rev. E \textbf{60}, 2721 (1999).


\bibitem{bioreview}A. Alemany, M. Ribezzi, and F. Ritort, AIP Conf. Proc. \textbf{1332}, 96 (2011).
\bibitem{molecule1} J. Liphardt, S. Dumont, S. B. Smith, I. Tinoco Jr., and C. Bustamante, Science \textbf{296}, 1832 (2002).
\bibitem{molecule2}D. Collin, F. Ritort, C. Jarzynski, S. B. Smith, I. Tinoco Jr., and C. Bustamante, Nature \textbf{437}, 231 (2005).

\bibitem{batalhao}T. B. Batalh\~{a}o, A. M. Souza, L. Mazzola, R. Auccaise, R. S. Sarthour, I. S. Oliveira, J. Goold, G. De Chiara, M. Paternostro, and R. M. Serra, Phys. Rev. Lett. \textbf{113},
140601 (2014).
\bibitem{cerisola}Federico Cerisola, Yair Margalit, Shimon Machluf, Augusto J Roncaglia, Juan Pablo Paz, and Ron Folman, 
%“Using a quantum work meter to test non-equilibrium fluctuation theorems,” 
Nat. Commun. {\bf 8}, 1241 (2017). 

\bibitem{pekola}O.-P. Saira, Y. Yoon, T. Tanttu, M. M\"ott\"onen, D. V. Averin, and J. P. Pekola, Phys. Rev. Lett. \textbf{109}, 180601 (2012).
\bibitem{pekola2}J. V. Koski and J. P. Pekola, in: Binder F., Correa L., Gogolin C., Anders J., and Adesso G. (eds), {\em Thermodynamics in the Quantum Regime}, Fundamental Theories of Physics, vol 195. Springer, Cham%arXiv:1805.10667 (2018).


%Extracting quantum work statistics and fluctuation theorems by single qubit interferometry
\bibitem{Dorner} R. Dorner, S. R. Clark, L. Heaney, R. Fazio, J. Goold, and V. Vedral, Phys. Rev. Lett. {\bf 110}, 230601 (2013).
%Measuring the characteristic function of the work distribution
\bibitem{Mazzola} L. Mazzola, G. De Chiara, and M. Paternostro, Phys. Rev. Lett. {\bf 110}, 230602 (2013).
%Lochschmidt echo in usdden and slow quenches
\bibitem{SquidLuttinger} B. D\'ora, F. Pollmann, J. Fort\'agh, and G. Zar\'and, Phys. Rev. Lett. \textbf{111}, 046402 (2013).

\bibitem{Dorosz}S. Dorosz, T. Platini, and D. Karevski, Phys. Rev. E \textbf{77}, 051120 (2008).
\bibitem{Silva}A. Silva, Phys. Rev. Lett. \textbf{101}, 120603 (2008).
\bibitem{Smacchia}P. Smacchia and A. Silva, Phys. Rev. E {\bf 88}, 042109 (2013).


\bibitem{Dora}B. D\'ora, \'A. B\'acsi, and G. Zar\'and, Phys. Rev. B \textbf{86}, 161109(R) (2012).


\bibitem{Gambassi}A. Gambassi and A. Silva, Phys. Rev. Lett. \textbf{109}, 250602 (2012).
\bibitem{Spyros}S. Sotiriadis, A. Gambassi, and A. Silva, Phys. Rev. E {\bf 87}, 052129 (2013).
\bibitem{Palmai}T. Palmai and S. Sotiriadis, Phys. Rev. E {\bf 90}, 052102 (2014).
\bibitem{Yi1}J. Yi, P. Talkner, and M. Campisi, {\it Nonequilibrium work statistics of an Aharonov-Bohm flux}, Phys. Rev. E \textbf{84}, 011138 (2011).
\bibitem{Yi2}J. Yi, Y. W. Kim, and P. Talkner, Phys. Rev. E \textbf{85}, 051107 (2012).
\bibitem{Perfetto}G. Perfetto, L. Piroli, and A. Gambassi, Phys. Rev. E \textbf{100}, 032114 (2019).

\bibitem{Quan2} Z. Fei, H. T. Quan, and Fei Liu, Phys. Rev. E \textbf{98}, 012132 (2018).
\bibitem{Quan1} Z. Fei and H.T. Quan, Phys. Rev. Lett. \textbf{124}, 240603 (2020).

\bibitem{wilkinsondiff1}M. Wilkinson, J. Phys. A: Math. Gen. \textbf{21}, 4021 (1988).

\bibitem{wilkinson3}M. Wilkinson and E. J. Austin, J. Phys. A: Math. Gen. \textbf{28}, 2277 (1995).
\bibitem{wilkinson4}M. Wilkinson, Phys. Rev. A \textbf{41}, 4645 (1990).
\bibitem{wilkinson7}M. Wilkinson and E. J. Austin, J. Phys. A: Math. Gen. \textbf{23}, L957 (1990).

\bibitem{Kravtsov1} M. A. Skvortsov, D. M. Basko, and V. E. Kravtsov, Pis’ma v ZhETF \textbf{80}, 60 (2004).
\bibitem{Kravtsov2} A. Ossipov, D. M. Basko, and V. E. Kravtsov, Eur. Phys. J. B \textbf{42}, 457–460 (2004).
\bibitem{Kravtsov3} D. M. Basko, M. A. Skvortsov, and V. E. Kravtsov, Phys. Rev. Lett. \textbf{90}, 096801 (2003).




\bibitem{Garcia-Mata}I. Garc\'ia-Mata, A. J. Roncaglia, and D. A. Wisniacki, Phys. Rev. E {\bf 95}, 050102(R) (2017).
\bibitem{Lobejko}M. \L{}obejko, J. \L{}uczka, and P. Talkner, 
%“Work distributions for random sudden quantum quenches,”
Phys. Rev. E \textbf{95}, 052137 (2017).
\bibitem{Arrais1}E. G. Arrais, D. A. Wisniacki, L. C. C\'eleri, N. G. de Almeida, A. J. Roncaglia, and F. Toscano, Phys. Rev. E \textbf{98}, 012106 (2018).
\bibitem{Arrais2} E. G. Arrais, D. A. Wisniacki, A. J. Roncaglia, and F. Toscano, Phys. Rev. E \textbf{100}, 052136 (2019).
\bibitem{delCampo1}A. Chenu, I. L. Egusquiza, J. Molina-Vilaplana, and A. del Campo, Sci. Rep. {\bf 8}, 12634 (2018).
\bibitem{delCampo2}A. Chenu, J. Molina-Vilaplana, and A. del Campo, Quantum \textbf{3}, 127 (2019).

%\bibitem{Marino}J. Marino and A. Silva, Phys. Rev. B {\bf 89}, 024303 (2014).


\bibitem{Grabarits1} I. Lovas, A. Grabarits, M. Kormos, and G. Zar\'and
Phys. Rev. Research \textbf{2}, 023224 (2020). 
\bibitem{Grabarits2} A. Grabarits, M. Kormos, I. Lovas, and G. Zar\'and, arXiv:2107.10245.



\bibitem{Kastnerreview} M. A. Kastner, Rev. Mod. Phys. \textbf{64}, 849 (1992).
\bibitem{Coulomb_book} \emph{Single Charge Tunneling: Coulomb Blockade Phenomena in Nanostructures}, eds. H. Grabert and M. H. Devoret (Plenum Press, New York, 1992)


\bibitem{calorimetry1}
S. Gasparinetti, K. L. Viisanen, O. P. Saira, T. Faivre, M. Arzeo, 
M. Meschke, and J.P. Pekola, Phys. Rev. Applied \textbf{3}, 014007 (2015).
%\bibitem{calorimetry2}E. D. Walsh, D. K. Efetov, G.-H. Lee, M. Heuck, J. Crossno, T. A. Ohki, P. Kim, D. Englund, and K. C. Fong, Phys. Rev. Applied \textbf{8}, 024022 (2017).

\bibitem{footnoteA} The coupling between the system and its thermal environment is assumed to be weak enough so that energy transfer between them is negligible during the quench \cite{workreview}.


\bibitem{matrixreview}C. W. J. Beenakker, Rev. Mod. Phys. \textbf{69}, 731 (1997).

\bibitem{Cardy} J. Cardy, Scaling and Renormalization in Statistical
Physics, Cambridge University Press (1996).
\bibitem{Sachdev} S. Sachdev, Quantum Phase Transitions, Cambridge University Press (2001).

\bibitem{Gasenzer} C.-M. Schmied, A. N. Mikheev, and T. Gasenzer, Int. J. Mod. Phys. {\bf 34}, 29 (2019).
\bibitem{boseuniversality} T. Langen, T. Gasenzer, and J. Schmiedmayer,
 J. Stat. Mech. 2016, 064009 (2016).
\bibitem{boseuniversality2} S. Erne, R. B\"ucker, T. Gasenzer, J. Berges, and J. Schmiedmayer, Nature \textbf{563}, 225 (2018). 
\bibitem{hydro1} C.  Cao,   E.  Elliott,   J.  Joseph,   H.  Wu,   J.  Petricka, T.  Sch\"afer,  and  J.  E.  Thomas, Science \textbf{331}, 58 (2011).
\bibitem{hydro2} A. Sommer, M. Ku, G. Roati, and M. W. Zwierlein, Nature \textbf{472}, 201 (2011).






\bibitem{mesobook}Y. Imry, Introduction to Mesoscopic Physics, (Oxford University Press, Oxford, 2002).
\bibitem{wilkinson2}P. N. Walker, M. J. S\'anchez, and M. Wilkinson, J. Math. Phys. \textbf{37}, 5019 (1996).
\bibitem{RMreview}T. Guhr, A. M\"uller-Groeling, and H. A. Weidenm\"uller, Phys. Rep. \textbf{299}, 189 (1998).
\bibitem{wilkinson6} E. J. Austin and M. Wilkinson, Nonlinearity \textbf{5}, 1137 (1992).



\bibitem{LZ1}L. Landau, Phys. Z. Sowj. \textbf{2}, 46 (1932).
\bibitem{LZ2}C. Zener, Proc. R. Soc. \textbf{137}, 696 (1932).
\bibitem{wilkinson5}M. Wilkinson, J. Phys. A: Math. Gen. \textbf{22}, 2795 (1989).
\bibitem{FeiQuan} Z. Fei and H.T. Quan, Phys. Rev. Research \textbf{1}, 033175 (2019).

\bibitem{Anderson} P. W. Anderson, 
% "Infrared Catastrophe in Fermi Gases with Local Scattering Potentials". 
Phys. Rev. Lett.  Review Letters. {\bf 18},  1049 (1967).


 \bibitem{footnoteB}We used typically $\sim 30000$ samples to generate the work statistics.
 
%\bibitem{JarzynskiPRE1993}  
%%Energy diffusion in a chaotic adiabatic billiard gas
%C. Jarzynski, Phys. Rev. E {\bf 48}, 4340 (1993).

\bibitem{MarkovChain}J. R. Norris, \emph{Markov Chains} (Cambridge University Press, Cambridge, 1998).
\bibitem{LargeIntSystem}Herbert Spohn, \emph{Large Scale Dynamics of Interacting Particles} (Springer, Berlin, Heidelberg, 1991).
\bibitem{StochIntSystem}Thomas M. Liggett, \emph{Stochastic Interacting Systems: Contact, Voter and Exclusion Processes} (Springer, Berlin, Heidelberg, 1999)
\bibitem{ASEP}B. Derrida, Phys.Rep. \textbf{301}, 65 (1998).
%\bibitem{CohenPRL2013} I. Tikhonenkov, A. Vardi, J. R. Anglin, and D. Cohen
%Phys. Rev. Lett. {\bf 110}, 050401 (2013).

\bibitem{vonDelft_KondoBox_paper}W. B. Thimm, J. Kroha, and J. von Delft, Phys. Rev. Lett. \textbf{82}, 2143 (1999). 


\bibitem{wilkinson1} M. Wilkinson and E. J. Austin, Phys. Rev A \textbf{47}, 2601 (1993).





%\bibitem{NVC}J. Klatzow, J. N. Becker, P. M. Ledingham, C. Weinzetl, K. T. Kaczmarek, D. J. Saunders,
%J. Nunn, I. A. Walmsley, R. Uzdin, and E. Poem, 
%%{\it Experimental Demonstration of Quantum Effects in the Operation of Microscopic Heat Engines}, 
%Phys. Rev. Lett. {\bf 122}, 110601 (2019). (NVC)
%\bibitem{CampisiOTOCM. Campisi and J. Goold, {\it Thermodynamics of quantum information scrambling}, Phys. Rev. E \textbf{95}, %062127 (2017). (OTOC)



\end{references}
\end{document}